\documentclass[article, onecolumn,showpacs,floats,floatfix,nofootinbib, usenatbib]{mn2e}

\usepackage{amsbsy}
\usepackage{amssymb}
\usepackage{amsmath}
\input{epsf}
\usepackage{graphicx}
\usepackage{color}

\usepackage{amssymb}
\usepackage{bm}
\def\spose#1{\hbox to 0pt{#1\hss}}

\def\lta{\mathrel{\spose{\lower 3pt\hbox{$\mathchar"218$}}
     \raise 2.0pt\hbox{$\mathchar"13C$}}}
\def\gta{\mathrel{\spose{\lower 3pt\hbox{$\mathchar"218$}}
     \raise 2.0pt\hbox{$\mathchar"13E$}}}

\newcommand{\p}{\mathrm{p}}

\newcommand{\n}{\mathrm{n}}
\newcommand{\s}{\mathrm{s}}

\begin{document}

\title[Dynamical tides in neutron stars]{Dynamical tides in neutron stars: The impact of the crust}

\author[Passamonti, Andersson and Pnigouras]
{A. Passamonti$^1$, N. Andersson$^2$ and P. Pnigouras$^{2,3,4}$ \\ \\
$^1$ Via Greve 10, 00146, Roma, Italy\\
$^2$ School of Mathematics and STAG Research Centre, University of Southampton, Southampton SO17 1BJ, UK\\
$^3$ Dipartimento di Fisica, ``Sapienza" Universit{\`a} di Roma \& Sezione INFN Roma1, Piazzale Aldo Moro 2, 00185 Roma, Italy\\
$^4$ Department of Physics, Aristotle University of Thessaloniki, 54124 Thessaloniki, Greece}

\maketitle

\date{\today}

\begin{abstract}
We consider the dynamical tidal response of a neutron star in an inspiralling binary, focussing on the impact of the star's elastic crust. Within the context of Newtonian gravity, we add the elastic aspects to the theoretical formulation of the problem and quantify the dynamical excitation of different classes of oscillation modes. The results demonstrate the expectation that the fundamental mode dominates the tidal response and show how the usual tidal deformability (and the Love number) emerge in the static limit. In addition, we consider to what extent the different modes may be excited to a level where the breaking strain of the crust would be exceeded (locally). The results show that the fundamental mode may fracture the crust during the late stages of inspiral. 
       This is also the case for the first gravity mode, which reaches the breaking threshold in strongly stratified stars. 
       In our models with a fluid ocean,  interface modes associated with the crust-ocean transition may also induce crust fracture. If this happens it does so earlier in the inspiral, at a lower orbital frequency. 
\end{abstract}

\begin{keywords}
stars: neutron, neutron star mergers, gravitational waves
\end{keywords}

%%%%%%%%
\section{Introduction and scope}

The breakthrough detections of gravitational waves from binary neutron star inspirals \citep{ligo1,love,ligo2} have led to renewed focus on the elusive neutron star equation of state. The problem has a number of complicating aspects---both relating to the observational data and the theoretical underpinning---but the essential question is quite simple: To what extent can we use observations to constrain the state and composition of matter under the extreme conditions that neutron stars represent?

Much of the recent focus has been on the neutron star tidal deformability, essentially the extent to which the tidal interaction with a binary companion deforms the neutron star fluid. This is a useful measure as it can be extracted from (or at least, constrained by) the gravitational-wave signal  \citep{hind,hind0}. Notably, the celebrated GW170817 event has led to a constraint on a suitable weighted average  tidal deformability,  corresponding (roughly) to a neutron star radius in the range 10-13~km \citep{love} (the result is somewhat model dependent). Morever, as this radius range agrees well with the  constraints obtained from the NICER observations of PSR J0030+0451 \citep{colem,riley} a consistent picture is beginning to emerge.

The deformability (often expressed in terms of the dimensionless Love number, $k_l$) represents the static contribution to the neutron star's tidal response. In addition, there is a dynamical tide. This is traditionally represented by the excitation of the different oscillation modes of the star. The resonance problem was first considered some time ago \citep{l94,reis,ks,ah}, but the issue is back in focus following the suggestion that the (fundamental) f-mode of the star may be excited to a relevant level, even though it may not reach resonance during the inspiral \citep{hind1,stein,ap20b}. The associated effect on the inspiral signal is weak, but its inclusion has been demonstrated to improve  waveform models. 

When we turn to dynamical features of the tide, we need to be mindful of the fact that a neutron star interior is a little bit more complicated than a prescribed pressure-density relation. Nuclei in the lower density region are expected to freeze to form the so-called crust, neutrons and protons likely form superfluid/superconducting condensates at high densities, there may be phase transitions to states with net strangeness (involving hyperons or deconfined quarks) and so on. These issues are important because the tidal response can be expressed as a sum over the (presumably complete set of) stellar oscillation modes \citep{l94,reis,ks} and the additional features of the interior physics may bring new modes into play and shift existing ones. On the one hand, this  makes the problem more complicated. On the other hand, it raises the question of whether the additional features may be observable. Even if this is not expected, one should take the care to demonstrate it (as there may be surprises!). A useful recent step in this direction \citep{ap20a} demonstrates that variations in the composition of the neutron star core affects the dynamical 
tide at the few percent level. This discussion also provides the (conceptually important) link between the static tide and the star's oscillation modes. Another important step was taken by \citet{wein} who considered the impact of mode resonances in a superfluid neutron star core, while \citet{pois} recently revisited the problem of (slowly) rotating stars, demonstrating that this issue also requires further thought. 

How do we do better? In principle, we know that the problem requires a fully relativistic analysis (one of the most detailed  seismology models to date was developed by \citet{christ}). However, the framework required to quantify the role of tidal resonances in a relativistic star has not yet been developed. There are technical issues to consider, including the fact that the modes of the system---which is now dissipative as gravitational waves are emitted---can not be (formally) complete. Quantitatively, this may not be important. Conceptually, it is a hurdle. While we ponder this issue, we may progress using phenomenological Newtonian models. In this direction, perhaps the most ``complete'' effort is provided by the results of \citet{pass} (see also \citet{hask}), which represent a star with an elastic crust penetrated by superfluid neutrons as well as a superfluid core (taking into account the entrainment effect in both regions) and an ocean. A reasonable aim then would be to consider the tidal response of such a star. This involves two steps. First, we need to develop the mode-coupling framework for a star with an elastic crust and ocean, and ensure that the mode calculation is sufficiently precise that we can quantify the (presumably) weak tidal excitation of (say) the crustal shear modes, the ocean g-modes and the interface modes \citep{mcd,mcd88}. The second step involves adding the superfluid components. With this paper, we take the first of these steps. We also consider the issue of tidally induced crust fracturing \citep{tsang,bonga}---to what extent resonant oscillation modes may reach the amplitude required to exceed the breaking strain at some point in the crust. 

%%%%%%
\section{Formulating the problem}

In order to set the stage for the analysis, let us remind ourselves of the context. The tide raised by a binary companion (treated as a point particle, which should be a good enough approximation for our purposes) induces a linear response in the primary. 
If we want to quantify the fluid response to this external agent---in essence, model the dynamical tide---we need to solve the linearised fluid equations in Newtonian gravity. As we are dealing with an (at least partly) elastic body, it is natural to do this in the framework of Lagrangian perturbation theory \citep{fs78}. In fact, many of the formalities have already been dealt with by \citet{hask} and \citet{pass}.

\subsection{The perturbation equations}

Assuming that the star is non-rotating, which makes sense on astrophysical grounds as the two partners in a binary  would have had plenty of time to spin down before the system enters the sensitivity band of a ground-based gravitational-wave detector (although, see the recent analysis of \citet{pois} for effects associated with spinning stars), we first of all have the  perturbed continuity equation 
\begin{equation}
\partial_t \left( \Delta \rho + \rho \nabla_i  \xi^i \right)  = 0  \, ,
\label{continuity}
\end{equation}
where $\xi^i$ is the  displacement vector associated with a Lagrangian perturbation (noting that, following \citet{fs78} we express vector components in a coordinate basis)
\begin{equation}
\Delta  = \delta + \mathcal L_\xi  
\end{equation}
(with $\delta$ the corresponding Eulerian variation and $\mathcal L_\xi$ the Lie derivative along $\xi^i$), such that the perturbed velocity is given by  
\begin{equation}
\Delta v^i = 
\delta  v^i = \partial_t  \xi^i \, .
\label{delv}
\end{equation}
Provided the background is in hydrostatic equilibrium, the perturbed Euler equation is
\begin{equation}
\partial_t^2  \xi_i +{1\over \rho}  \nabla_i \delta p - {1\over \rho^2} \delta \rho \nabla_i p + \nabla_i \delta \Phi = - \nabla_i \chi
\label{eul2}
\end{equation}
and we also  have the Poisson equation for the gravitational potential 
\begin{equation}
\nabla^2 \delta \Phi = 4\pi G \delta \rho \, ,
\label{pois1}
\end{equation}
while the tidal potential, $\chi$, due to the presence of the binary partner (which generates the fluid perturbation), is a solution to $\nabla^2 \chi = 0$.

As we progress, it is useful to note that \eqref{continuity} leads to
\begin{equation}
 \Delta \rho + \rho \nabla_i  \xi^i = 0 \quad \Longrightarrow \quad \delta \rho = - \nabla_i \left( \rho \xi^i \right) \, , 
 \label{drho}
\end{equation}
which allows us to write the Poisson equation as
\begin{equation}
\nabla_i \left( g^{ij} \nabla_j \delta \Phi + 4\pi G \rho \xi^i \right) = 0 \, .
\label{gpalt}
\end{equation}
This will be useful later, as we can integrate to get
\begin{equation}
\nabla_i \delta \Phi = - 4\pi G  \rho \xi_i + S_i
\end{equation}
for some vector $S^i$ such that $\nabla_i S^i = 0 $. It is easy to see that, if $\rho\to 0 $ at the star's surface, then we must have
\begin{equation}
S_i = \partial_i \delta \Phi \ , \quad \mbox{at}\ r=R    \, ,
\end{equation}
and we know that this should not vanish. 
Moreover, we know that the solution has to match to the vacuum exterior, which means that (again, as long as\footnote{Later, when we consider the contribution from the elastic crust, we  assume that the crust is covered by a shallow fluid ocean. This makes sense on physical grounds, which is convenient as the surface boundary conditions then remain those of a perturbed fluid star.} the density  vanishes as $r\to R$)
\begin{equation}
{d \over dr} \delta  \Phi_l + {l+1 \over r} \delta  \Phi_l =0 \quad \quad \mbox{at} \quad r=R \, ,
\label{surfcon}
\end{equation}
where we have expanded $\delta \Phi$ in spherical harmonics and focussed  on a single $l$ multipole.
That is, we must have 
\begin{equation}
    \hat n^i S_i =  \partial_r \delta \Phi =  \left( \partial_r \delta \Phi_l \right) Y_l^m  = - {l+1\over r} \delta \Phi \quad\quad \mbox{at} \quad r=R
\end{equation}
where $\hat n_i$ is the normal to the star's spherical surface (i.e. parallel to the radial basis vector).
We will have reason to recall this result later.

%%%%%%
\subsection{Mode orthogonality}

We now want to express  the driven tidal response of the  star---the solution to \eqref{eul2}---in terms of a set of normal modes \citep{l94,reis,ks,ap20a}, corresponding to solutions $\xi^i_n$ (with $n$ labelling the modes). Letting the (real) mode frequency be $\omega_n$,  each individual mode then satisfies the homogeneous version of \eqref{eul2}, with $\chi=0$.

In order to provide a complete picture, it is useful to first consider a barotropic model for which we have (see \citet{kirsty} for a similar analysis)
\begin{equation}
\delta p  = n \delta \mu = \rho \delta \tilde \mu \, , \label{eq:dp}
\end{equation}
where $\mu= m_\mathrm{B} \tilde \mu$ is the chemical potential and $\rho = m_\mathrm{B} n$, with $m_\mathrm{B}$ the mass of each baryon and $n$ the baryon number density. 
This means that. for a background in hydrostatic equilibrium, we have 
\begin{equation}
{1\over \rho}  \nabla_i \delta p - {1\over \rho^2} \delta \rho \nabla_i p 
= \nabla_i \delta \tilde \mu \, .
\end{equation}
As a result, we have
\begin{equation}
\partial_t^2  \xi_i  +  \nabla_i \left( \delta \Phi + \delta \tilde \mu \right) = 0 \, ,
\label{eulmu}
\end{equation}
or, if we take the baryon number density to be the primary matter variable;
\begin{equation}
\partial_t^2  \xi_i  +  \nabla_i \left[ \delta \Phi + \left({\partial \tilde \mu \over \partial n} \right) \delta n  \right] = \partial_t^2  \xi_i  +  \nabla_i \left[ \delta \Phi - \left({\partial \tilde \mu \over \partial n} \right)\nabla_j \left( n \xi^j\right) \right] = 0 \, .
\end{equation}

Following \citet{fs78} we now write the Euler equation as
\begin{equation}
-\omega_n^2 A \xi_n^i + C \xi_n^i = 0 \, ,
\end{equation}
where 
\begin{equation}
    A=\rho 
\end{equation}
and
\begin{equation}
 C \xi_n^i = \rho \nabla^i \left[ \delta \Phi -  \left({\partial \tilde \mu \over \partial n} \right)\nabla_j \left( n \xi_n^j\right) \right]
\end{equation}
and introduce the inner product 
\begin{equation}
\langle \eta , \rho  \xi \rangle = \int \rho  \eta_i^* \xi^i dV \, ,
\end{equation}
where the asterisk indicates the complex conjugate and $\eta^i$ is another  solution to the perturbation equations. 
It is then fairly easy to show that mode solutions are orthogonal if the mode frequencies are real. 
As we will need to extend the proof of this to the elastic case, let us have a look at the argument---taking the opportunity to keep careful track of ``surface terms'' that come into play if we allow internal phase transitions. 

First of all, it is obvious from the definition of the inner product that
\begin{equation}
\langle\eta ,A \xi\rangle = \langle\xi,A \eta\rangle^*   \ . \label{sym1}
\end{equation}
The argument for the $C$ operator is a bit more involved. We need to use integration by parts to show that
\begin{multline}
    \int n \eta^{i*} \nabla_i \left[ \left( {\partial \tilde{\mu} \over 
\partial n} \right) \nabla_j (n\xi^j)  \right] dV  \\
 = \int n \left( {\partial \tilde{\mu} \over 
\partial n} \right) \left[ \eta^{i*} \nabla_j (n\xi^j)   -  \xi^i   \nabla_j  (n\eta^{j*})  \right] \hat n_i dS + 
\int  n\xi^j \nabla_j \left[ \left( {\partial \tilde{\mu} \over 
\partial n} \right)  \nabla_i  (n\eta^{i*}) \right] dV
\\
= 
\int n \xi^i \nabla_i \left[ \left( {\partial \tilde{\mu} \over \partial n} \right) \nabla_j (n\eta^{j*})  \right] dV \, ,
\label{muInt}
\end{multline}
where $\hat n_i$ is the (outwards pointing) normal to the surface of the volume we are integrating over. The last equality holds if can ignore the contribution from the surface term. This would certainly be the case if we integrate over the entire star, as long as $n\to 0$ at the surface, but there may be situations where one would need to be more careful.

For example, as we may want to consider models with internal phase transitions---e.g. at the crust-core interface---it is important to establish what happens if such a transition is associated with a discontinuity in the density. It is easy to see that \eqref{muInt} implies that, if there is an internal density jump, at $r=\bar R$, say, we need the surface term
\begin{equation}
     n \left( {\partial \tilde{\mu} \over 
\partial n} \right) \left[ \eta^{i*} \nabla_j (n\xi^j)   -  \xi^i   \nabla_j  (n\eta^{j*})  \right] \hat n_i
\end{equation}
to be continuous. 
To see that this should be the case, use (again for a barotrope) $d p = n  d  \mu$ to get 
\begin{multline}
      {1\over m_B}\left( {\partial p \over 
\partial n} \right) \left[ \eta^{i*} \nabla_j (n\xi^j)   -  \xi^i   \nabla_j  (n\eta^{j*})  \right] \hat n_i = -  {1\over m_B} \left( {\partial p \over 
\partial n} \right) \left[ \eta^{i*} \delta_\xi n   -  \xi^i   \delta_\eta n^*  \right] \hat n_i  \\
 = -  {1\over m_B}  \left[ \eta^{i*} \delta_\xi p   -  \xi^i   \delta_\eta p^*  \right] \hat n_i \, , \label{eq:eq-dis}
\end{multline}
where $\delta_\xi$ and $\delta_\eta$ distinguishes the Eulerian perturbations associated with $\xi^i$ and $\eta^i$, respectively. Continuity across a surface requires (i) the radial component of the displacement, and (ii) the Lagrangian pressure variation, to be continuous. 
The last term of equation (\ref{eq:eq-dis}) can be rewritten as  
\begin{equation}
\left[ \eta^{i*} \delta_\xi p   -  \xi^i   \delta_\eta p^*  \right] \hat n_i  =   \left[ \eta^{i*} \Delta_\xi p   -  \xi^i   \Delta_\eta p^* - \left( \eta^{i*} \xi^j \nabla_j p   -  \xi^i   \eta^{*j} \nabla_j p \right) \right] \hat n_i  \, .
\end{equation}
For a spherical star, this expression is clearly continuous as the background pressure depends only on the radial coordinate.  

Turning to the gravitational potential, we have 
\begin{multline}
\int \rho \eta^{i*} \nabla_i \delta_\xi \Phi dV   = \int \left[ \rho \eta^{i*} \delta_\xi \Phi \right] \hat n_i dS - \int  \delta_\xi \Phi \nabla_i ( \rho \eta^{i*}) dV \\
    = 
  \int \left[ \rho \left(\eta^{i*} \delta_\xi \Phi - \xi^i \delta _\eta \Phi^*\right)  + { 1 \over 4 \pi G}  g^{ij} \left( \delta_\xi \Phi  \nabla_j \delta_\eta \Phi^* - \delta_\eta \Phi^*   \nabla_j \delta_\xi \Phi \right) \right] \hat n_i dS \\ 
  + \int   \rho \xi^i \nabla_i  \delta_\eta \Phi^* dV
   = 
\int   \rho\xi^i \nabla_i  \delta_\eta \Phi^* dV  \ ,
\label{PhiIntegral}
\end{multline}
where the last identity (again) holds as long as we may ignore the surface terms.

In particular, at the surface of the star we need to ensure that
\begin{equation}
     \rho \left(\eta^{i*} \delta_\xi \Phi - \xi^i \delta _\eta \Phi^*\right)  + { 1 \over 4 \pi  G}  g^{ij} \left( \delta_\xi \Phi  \nabla_j \delta_\eta \Phi^* - \delta_\eta \Phi^*   \nabla_j \delta_\xi \Phi \right) = 0 \, .
     \label{surfpot}
\end{equation}
The first term (obviously) vanishes as long as $\rho \to 0$ at the surface. Moreover, noting that the matching to the exterior potential requires \eqref{surfcon} to hold, 
we see that the second term vanishes, as well. Specifically, we have (for each multipole)
\begin{equation}
 \delta_\xi \Phi \partial_r  \delta_\eta \Phi^* - \delta_\eta \Phi^*   \partial_r  \delta_\xi \Phi =   - \frac{l+1}{R} \left( \delta_\eta \Phi^*  \delta_\xi \Phi - \delta_\xi \Phi \delta_\eta \Phi^* \right)_{r=R} = 0 \, .
\end{equation}

The conditions that apply when the density does not vanish at the surface, can be inferred from the results for an internal density jump. 
Again, assume that there is a density jump at an internal point, $r=\bar R$, such that (in a small neighbourhood of the phase transition)
\begin{equation}
    \rho(r) = \left\{ \begin{array}{ll} \rho^-\ , \quad r<\bar R \\ \rho^+ \ , \quad r >\bar R  \end{array} \right.
\end{equation}
and integrate \eqref{gpalt} over a small volume (with height $=2\epsilon$) across the (spherical) surface to get
\begin{equation}
        \left[ \partial_r \delta \Phi + 4\pi G \rho \xi_r \right]^{\bar R + \epsilon}_{\bar R - \epsilon} = 2\epsilon S
\end{equation}
for some constant $S$. Then let $\epsilon \to 0$ to get the junction condition
\begin{equation}
    \partial_r \delta \Phi^+ + 4\pi G \rho^+ \xi_r = \partial_r \delta \Phi^- + 4\pi G \rho^- \xi_r \, ,
    \label{junc1}
\end{equation}
where we have used the fact that the radial component of the displacement must be continuous. What does this mean for the surface terms in \eqref{PhiIntegral}? Well, we integrate up to $\bar R$ and then continue on the other side. The surface terms associated with the interface will vanish as long as 
\begin{equation}
 \hat n_i \left[  \rho \left(\eta^{i*} \delta_\xi \Phi - \xi^i \delta _\eta \Phi^*\right)  + { 1 \over 4 \pi  G}  g^{ij} \left( \delta_\xi \Phi  \nabla_j \delta_\eta \Phi^* - \delta_\eta \Phi^*   \nabla_j \delta_\xi \Phi \right) \right]^{\bar R+\epsilon}_{\bar R-\epsilon} \to 0 \quad \mbox{as}\ \epsilon \to 0  \, .
   \end{equation}
 We can rewrite this as
\begin{equation}
  \left[ \delta_\xi \Phi\left( \partial^r \delta_\eta \Phi + 4\pi G  \rho \eta^{r} \right)^* - \delta _\eta \Phi^* \left( \partial^r \delta _\xi \delta \Phi +  4 \pi  G \rho \xi^r  \right) \right]^{\bar R+\epsilon}_{\bar R-\epsilon} \to 0 \quad \mbox{as}\ \epsilon \to 0  \, .
\end{equation}
Combining the fact that the perturbed gravitational potential is continuous with \eqref{junc1}, we see that these contributions must vanish. It is also easy to adjust the argument to arrive at the standard result for the gravitational potential matching at a finite-density surface. In effect, we have provided a (somewhat lengthy) demonstration that we do not have to worry about density discontinuities in the following. As long as the relevant junction conditions are respected, the usual orthogonality argument remains unchanged\footnote{The argument that discontinuities do not impact on the mode orthogonality as long as the relevant junction conditions are respected should hold in general. This is important for more realistic neutron star models, which may have phase transitions in the fluid core. A more realistic model will also have discontinuities in the crust shear modulus (associated with layers of nuclei with different atomic numbers). Our argument applies to these discontinuities, as well.}. 

At the end of the day, we have the expected result \citep{fs78}:
\begin{equation}
\langle\eta ,C \xi\rangle = \langle\xi,C \eta\rangle^*   \ . \label{sym2}
\end{equation}
This means that, if we consider two mode solutions (now letting $\xi_n$ be associated with frequency $\omega_n$ and $\xi_m$ be the solution associated with $\omega_m$) we have 
\begin{multline}
0 = \langle  \xi_m , -\omega_n^2 A \xi_n + C \xi_n \rangle = - \omega_n^2 \langle \xi_n, A  \xi_m\rangle^* + \langle \xi_n , C  \xi_m\rangle^* \\
=  - \omega_n^2 \langle \xi_n, A  \xi_m\rangle^*+ (\omega_m^*)^2  \langle \xi_n, A  \xi_m\rangle^* = \left[ (\omega_m^*)^2 -  \omega_n^2 \right]  \langle \xi_n, A  \xi_m\rangle^* \, .
\label{orto1}
\end{multline}
We see that, if the frequencies are distinct and real, the mode solutions must be orthogonal. That is, we may use
\begin{equation}
\langle \xi_n, A  \xi_m\rangle = \mathcal A_n^2\delta_{mn} \, ,
\label{norm}
\end{equation}
leaving the normalisation $ \mathcal A_n^2$ unspecified for the moment.

So far, the arguments are standard. The only aspect we have added relates to internal phase transitions. 
Before we move on let us, without detailed calculation, comment on the case of non-barotropic perturbations, relevant whenever we want to account for internal matter stratification. It is easy to argue that this case applies to neutron star tides as the relevant nuclear reactions are too slow to keep the perturbed matter in beta equilibrium \citep{ap20a}. In particular, it leads to the presence of gravity g-modes in the oscillation spectrum. 

In the limit of slow reactions we may describe the equation of state in terms of two parameters. The most natural choice is to complement the density $\rho$ with either a parameter representing the deviation from chemical equilibrium or the proton fraction $x_\p = \rho_\p /\rho$, the Lagrangian perturbation of which will vanish for slow reactions. However, the mode orthogonality can be established without assuming an explicit equation of state (keeping both the pressure and the density perturbations). The argument follows from  \citet{fs78}, and we arrive at 
\begin{equation}
 \int \eta^{i*} \left[ \nabla_i \delta_\xi p  -  {1\over \rho}  \delta_\xi \rho \nabla_i p\right]  dV %\\
 =  \int \left[ \eta^{i*} \delta_\xi p - \xi^i \delta_\eta p^* \right] \hat n_i dS +  \int\xi^i \left[  \nabla_i \delta_\eta p^* -  {1\over \rho}  \delta_\eta \rho^* \nabla_i p \right]dV \, .
 \label{stratcon}
\end{equation}
This is the result we need to prove the mode orthogonality for stratified stars. As the argument for the gravitational potential remains unchanged, the stated symmetry relation \eqref{sym2} will hold as long as the radial displacement and the pressure perturbation are both continuous (for a spherical surface).

%%%%%%%%%%
\subsection{Adding the crust}

We now take the ``natural'' next step towards a more realistic neutron star model, adding the elastic crust that is expected to reach from (close to) the surface up to about 60-70\% of the nuclear saturation density (roughly, the outer kilometer of the star). Assuming that the crust of the equilibrium star is ``relaxed''---not associated with strains, the elasticity impacts only on the perturbations. The argument is similar to that used in previous work on the tidal deformability and neutron star ``mountains'', see for example \citet{fabian}. The Euler equation then changes to
\begin{equation}
-\omega_n^2 A \xi^i + C \xi^i + E \xi^i = 0 \, ,
\end{equation}
 where
\begin{equation}
E \xi_i = \nabla^j \sigma_{ij} 
\end{equation}
with the elastic stress tensor given by
\begin{equation}
\sigma_{ij} = \check \mu \left( \nabla_i \xi_j + \nabla_j \xi_i\right) - \frac{2}{3} \check \mu g_{ij} \left( \nabla_k \xi^k \right) \, .\label{eq:sigij}
\end{equation}
Note that the shear modulus $\check \mu$  must not to be confused with the chemical potential.

Now, we need
\begin{equation}
\langle \eta^{i*}, E \xi_i \rangle = \int  \eta^{i*}  \nabla^j \sigma_{ij} dV 
= \int \left[ \eta^{i*} \sigma_{ij} \right]\hat n^j dS - \int \sigma_{ij} \nabla^j \eta^{i*} dV \, .
\end{equation}
The second term has two pieces:
\begin{equation}
 \int \check \mu \left( \nabla_i \xi_j + \nabla_j \xi_i \right) \nabla^j \eta^{i*}   dV = \int \check \mu \left( \nabla_i \eta^*_j+ \nabla_j \eta^*_i  \right)\nabla^j \xi^i dV
\end{equation}
and
\begin{equation}
\int \check \mu g_{ij} \left( \nabla^k \xi_k \right)  \nabla^j \eta^{i*} dV = \int \check \mu \left( \nabla^k \xi_k \right) \left( \nabla^j \eta^*_j \right) dV = \int \check \mu g_{ij} \left( \nabla^k \eta^*_k \right) \left( \nabla^j \xi^i \right)  dV \, .
\end{equation}
Introducing
\begin{equation}
\bar \sigma_{ij} = \check \mu \left( \nabla_i \eta^*_j + \nabla_j \eta^*_i\right) - \frac{2}{3} \check \mu g_{ij} \left( \nabla^k \eta^*_k \right) \, ,
\end{equation}
we see that 
\begin{multline}
\langle \eta^{i*}, E \xi_i \rangle 
= \int \left[ \eta^{i*} \sigma_{ij} \right]\hat n^j dS  - \int \bar \sigma_{ij} \nabla^j \xi^{i} dV =  \int \left[ \eta^{i*} \sigma_{ij} - \xi^i \bar \sigma_{ij} \right]\hat n^j dS  + \int \xi^{i}\nabla^j \bar \sigma_{ij}   dV \\
= \int \left[ \eta^{i*} \sigma_{ij} - \xi^i \bar \sigma_{ij} \right]\hat n^j dS + \langle \xi^i, E\eta^*_i\rangle \, .
\end{multline}

In order to deal with the surface terms we need to consider the traction conditions \citep{hask}. 
Hence, we impose the continuity of the
gravitational potential and the perpendicular and radial components of the traction vector
\begin{equation}
 t^i = - \left( g^{ij} \Delta p  - \sigma^{ij} \right) \hat n_j 
\end{equation}
noting that the traction reduces to the pressure perturbation in the fluid regions. 
In order to establish that the surface terms vanish at the crust-core transition, we combine the above result with the relevant term from \eqref{muInt}. This leads to the requirement that 
\begin{multline}
    - \left[ \eta^{i*} \delta_\xi p   -  \xi^i   \delta_\eta p^*  \right] \hat n_i + \left[ \eta^{i*} \sigma_{ij} - \xi^i \bar \sigma_{ij} \right]\hat n^j \\
    =  - \left[ \eta^{i*} \left( g_{ij}\Delta_\xi p   -   \sigma_{ij} \right) -   \xi^i  \left( g_{ij} \Delta_\eta p^* -  \bar \sigma_{ij} \right) -  g_{ij} ( \eta^{i*}   \xi^k   -  \xi^i \eta^{*k} ) \nabla_k p \right]\hat n^j 
\end{multline}
is continuous at the interface. We see that, given the traction conditions, this should always be the case for a spherical star. In essence, the addition of an elastic region does not  complicate the formal analysis.

%%%%%%%%%%%
\subsection{Expanding the tidal response}
\label{expand}

Finally, we are set to return to the tidal problem. This part of the analysis is straightforward given the properties we have established.  Perhaps not surprisingly, the results we need are identical to those from the fluid case, see, for example, \citet{ap20a}. In particular, it is clear that the oscillation modes of the fluid+elastic problem are orthogonal and that \eqref{norm} still holds. As in the fluid problem, we can use this fact to  rewrite the Euler equation
\begin{equation}
\rho \partial_t^2  \xi_i + C \xi_i + E\xi_i =  -\rho \nabla_i \chi
\label{eul3}
\end{equation}
as an equation for individual mode amplitudes. Introducing the mode expansion
\begin{equation}
 \xi^i = \sum_n a_n(t)   \xi_n^i  \, , \label{eq:xi_a}
\end{equation}
where the eigenfunctions  $\xi_n^i$  are time-independent (as they are obtained in the frequency domain), we easily arrive at
\begin{equation}
\ddot a_n + \omega_n^2 a_n = - {1\over  \mathcal A_n^2} \langle \xi_n, \rho \nabla \chi\rangle 
=  - {1\over  \mathcal A_n^2}\int \chi \delta \rho^*_n dV 
\label{eq:eqa}
\end{equation}
making use of the continuity equation and integrating by parts in the last step.

Expanding the tidal potential in spherical harmonics \citep{ap20a} we have
\begin{equation}
 \chi  = \sum_{l,m}   v_l r^l Y_{lm} \, ,
\label{fdef}
\end{equation}
where the $v_l$ coefficients will be given later. Assuming an adiabatic inspiral and working in the frequency domain (with time dependence $e^{i\omega t}$), the tidal driving only has support at frequency $\omega = m\Omega$. It is useful to keep this in mind. Anyway, we arrive at a set of driven  modes
with amplitude 
\begin{equation}
a_n =   {1 \over \omega_n^2 - \omega^2}  {Q_{n} \over \mathcal A_n^2}  v_{l} \, ,
\label{reason}
\end{equation}
where we have introduced the overlap integral
\begin{equation}
Q_{n} = - \int \delta \rho^*_n r^{l+2}  dr \, .
\label{overlap}
\end{equation}

As discussed by \citet{ap20a} the overlap integral can also be expressed in terms of the matching of the gravitational potential at the star's surface
\begin{equation}
4\pi G \int_0^R r^{l+2} \delta  \rho dr = - (2l+1) R^{l+1}  \delta  \Phi (R) \, ,
\end{equation}
so we have
\begin{equation}
Q_{n}  =   { 2l+1 \over 4\pi G } R^{l+1}  \delta  \Phi_n (R) =  I_n \, ,
\label{multip}
\end{equation}
where $I_n$ represents the contribution each of the star's oscillation modes makes to the mass multipole moment\footnote{The argument here is analogous to the mode-sum rule for the multipole moments discussed by \citet{reisb}. Our general arguments then demonstrate that this mode-sum is not affected by discontinuities, e.g. associated with phase transitions, as long as the relevant junction/traction conditions are respected in the mode calculation.}.

Finally, we  connect the mode expansion to the tidal deformability and the effective Love number by introducing a representation for the perturbed gravitational potential in terms of the mode eigenfunctions.
Expressing the displacement vector for each multipole as 
\begin{equation}
\xi^i =  \left( W(r)  {\nabla^i r  \over r} \right) Y_{lm}\ + V(r) \nabla^i Y_{lm}  \label{eq:xiexp} \, ,
\end{equation}
it follows from the $\theta$-component of \eqref{eul2} that
(since  $\Delta p = 0$ at the surface)
\begin{equation}
-\omega^2  V(R) -{ p' \over \rho}  {W(R) \over R}  + \delta \Phi(R) = -\chi(R)
\label{eulc} \, .
\end{equation} 
That is, we have
\begin{equation}
\delta \Phi(R) = - \chi(R) + \omega^2 V(R) - g  {W(R) \over R}
\end{equation}
and we arrive at an expression for the dynamical tidal response
\begin{equation}
k_l^\mathrm{eff} = {1\over 2} {\delta \Phi(R) \over \chi(R) } = -{1\over 2} + {1\over 2 v_l R^l} \left[ \omega^2 V(R) - g {W(R)\over R} \right] \, .
\end{equation}
Moreover, making contact with \eqref{reason} we see that 
\begin{equation}
\xi^i =  \sum_n {1 \over \omega_n^2 - \omega^2} {Q_n v_l \over \mathcal A_n^2} \xi^i_n \, ,
\end{equation}
which means that we have (simply adding a label, $n$, to indicate the individual mode eigenfunctions) 
\begin{equation}
k_l^\mathrm{eff} =  -{1\over 2} + {1\over 2 R^l} \sum_n {Q_n \over \mathcal A_n^2} {1 \over \omega_n^2 - \omega^2}\left[ \omega^2 V_n(R) - {GM_\star\over R^3} W_n(R)\right]
\label{keff1}
\end{equation}
with $M_\star$ being the primary's mass. Keeping the normalisation of the modes unspecified---which may be useful when we consider the numerical aspects of the problem---we have
\begin{equation}
\mathcal A_n^2 = \int_0^R \rho \left[ W_n^2 + l(l+1) V_n^2\right] dr \, .
\label{normal}
\end{equation}
Introducing the dimensionless frequency 
\begin{equation}
    \tilde \omega^2 = {\omega^2 R^3 \over GM_\star}
\end{equation}
as well as  the dimensionless overlap integral
\begin{equation}
\tilde Q_n = {Q_n \over M_\star R^l} \label{eq:dimQn} \, ,
\end{equation}
we get
\begin{equation}
k_l^\mathrm{eff} =  - {1\over 2} - {1\over 2 } \sum_n {\tilde Q_n  \over \tilde \omega_n^2 - \tilde \omega^2}
{  M_\star  W_n(R) \over \mathcal A^2_n} \left[ 1-  \tilde \omega^2 \left( {V_n \over W_n}\right)_R\right] \, .
\label{keff2}
\end{equation}
This is the final result, but
we can massage it a bit by recalling
 \eqref{multip} and using
\begin{equation}
\delta \Phi_n(R) =   \omega_n^2 V_n(R) - g  {W_n(R) \over R} \, .
\end{equation}
We have \citep{ap20a}
\begin{equation}
W_n(R) = - {4\pi\over 2l+1} {\tilde Q_n R^2} \left[ 1 - \tilde \omega_n^2 \left( {V_n\over W_n}\right)_R\right]^{-1}
\end{equation}
which, when combined with  \eqref{keff2}, leads to the final expression
\begin{equation}
k_l^\mathrm{eff} =  -{1\over 2} + {2\pi \over 2l+1} \sum_n {\tilde Q_n^2 \over \tilde \omega_n^2 - \tilde \omega^2} \left({ M_\star R^2 \over \mathcal A^2_n} \right) \left[ 1-  \tilde \omega^2 \left( {V_n \over W_n}\right)_R\right]  \left[ 1 - \tilde \omega_n^2 \left( {V_n\over W_n}\right)_R\right]^{-1} \, .
\end{equation}

In the low-frequency limit, we obtain a mode sum for the tidal Love number
\begin{equation}
k_l \approx -{1\over 2} + {2\pi \over 2l+1} \sum_n {\tilde Q_n^2 \over  \tilde \omega_n^2 } \left({ M_\star R^2 \over \mathcal A^2_n} \right) \left[ 1 - \tilde \omega_n^2 \left( {V_n\over W_n}\right)_R\right]^{-1} =  - {1\over 2} + \sum_n k_l^n \, .
\label{klfinal}
\end{equation}

The careful reader will note that the expressions for the effective Love number remain unchanged from \citet{ap20a}. This is to be expected, as long as we consider a neutron star with a fluid (ocean) surface. The results would change if we were to impose a direct transition from the elastic region to the exterior vacuum as we would then have to consider elastic terms in, for example, \eqref{eulc}. A fluid ocean is expected on physical grounds (even though it may be very shallow for mature neutron stars). Moreover, as the ocean may sustain its own (more or less distinct) set of oscillation modes (g-modes) it is relevant to include this aspect in the model.

%%%%%%%%
\section{The dynamical tide} \label{sec:results}

Having discussed the theoretical framework, let us move on to consider the numerical results for a sample of neutron star models. In order to facilitate a direct comparison, we have chosen to focus on the stratified polytropic models already considered by \citet{ap20a}.  The stellar models we consider then involve three distinct regions: the single fluid core, the elastic crust and a (shallow) fluid ocean.  In effect, there will be new features, like shear modes associated with the elasticity and interface modes linked to the core-crust and crust-ocean transitions \citep{mcd,mcd88}. We do not consider the impact of superfluidity in the core or the crust at this point, but plan to return to this problem later.

In the first instance, we are interested in the static and dynamical aspects of the tide, as represented (via the different oscillation modes) by the tidal deformability and the Love number $k_l^\mathrm{eff}$. As we anticipate the impact of the elasticity on already existing fluid modes to be slight and  the tidal excitation of, for example, the shear oscillations in the crust to be weak, we have to make sure our numerical approach is robust. We need to do better than the proof-of-principle analysis of \citet{ap20a}. Indeed, in order to reach the desired precision in the relevant eigenfunctions and the overlap integrals, we have to improve on the approach from \citet{pass}. The results we present in the following were obtained using a ``classic'' shoot-and-relax approach, where a shooting method was used to identify each oscillation mode and the precision was subsequently improved by a relaxation step. The fact that this implementation allows us to reliably extract the quantities we are interested in should be evident from the results we provide.

Specifically, we consider a simple polytropic equation of state,  $p=k \rho^{\Gamma}$,  with  $\Gamma = 2$. To account for the effects of stratification on the oscillation modes and the tide, we introduce an adiabatic index, $\Gamma_1$, for the perturbations: 
\begin{equation} 
\frac{\Delta \rho}{\rho} = \frac{1}{\Gamma_1} \frac{\Delta p}{ p} \, . \label{eq:G1}
\end{equation}
Following \citet{ap20a} we consider three models, a  barotropic star with $\Gamma_1 = 2$ and weakly and strongly stratified stars  with, respectively,  $\Gamma_1 = 2.05$ and 
$\Gamma_1 = 7/3$.  
For simplicity, we assume that this equation of state describes the entire star, from the core to the ocean. This may not be  particularly realistic, but as we are working in the framework of Newtonian gravity we have to settle for a somewhat phenomenological set-up.
From equation (\ref{eq:G1}) it follows that 
we can write the relation between the Eulerian perturbations of pressure and mass density 
\begin{equation}
\frac{\delta \rho}{\rho} = 
 \frac{1}{\Gamma_1} \frac{\delta p}{p} - A_i \xi^i \, ,
\end{equation}
where 
\begin{equation}
A_i =    \nabla_i \ln \rho - \frac{\nabla_i \ln p}{\Gamma_1}  
\end{equation}
is the Schwarzschild discriminant, which quantifies the presence of buoyancy and determines 
the properties of gravity modes. 

All numerical results presented in the following were obtained with the crust-core transition taken to be at $R_{\rm cc} = 0.9 R$ and  the crust-ocean interface at $R_{\rm co} = 0.999 R$. 
The elastic properties of the crust are described by the shear modulus $\check \mu$. Following \citet{douchin} we assume that the  specific shear modulus is nearly constant across the crust. This would mean taking
\begin{equation}
\frac{\check \mu}{\rho} \simeq 10^{16} \, \textrm{cm}^2/ \textrm{s}^{2}. 
\label{douch}
\end{equation}
but in practice we use the (further) simplified version from \citet{pass} and let
\begin{equation}
\check \mu =  \tilde \mu \, \rho  \,  \left( \frac{G M_{\star} }{ R} \right)   \label{eq:mu} \, ,
\end{equation}
where $\tilde \mu $ is a dimensionless parameter, which must not be confused with the chemical potential defined in equation (\ref{eq:dp}). For a star with $M_{\star}=1.4 M_{\odot}$ and $R=10\,$km 
we then find  $G M_{\star}/R=1.86 \times 10^{20} \rm cm^2/s^2$. Linking to \eqref{douch} we  assume the 
value  $\tilde \mu =10^{-4}$ for most of our examples, exploring other cases only for the interface modes.

%%%%%%%%%%%%
\subsection{The oscillation modes}
\label{sec:modes}

Even though it is fairly simple, the stellar model we consider  sustains many (more or less) distinguishable classes of oscillation modes:  the fundamental mode (f mode), the pressure modes (p modes), the gravity modes (g modes), 
the shear modes (s modes) and the interface modes (i modes) (see \citet{mcd} and \citet{mcd88} for detailed discussions).  
The f and p modes are present also in a fluid star, while the g modes appear in stratified models (in our case 
  when $\Gamma_1 \neq \Gamma$), see \citet{ap20a}.
  The presence of the crust impacts on these classes of modes and introduces, due to the elasticity, the shear modes. 
  Moreover,  any transition between different regions in the star (core-crust and crust-ocean) tends to be associated with a set of interface modes.

%-------------FIG. 1------------------------------------------%
\begin{figure}
\begin{center}
\includegraphics[height=70mm]{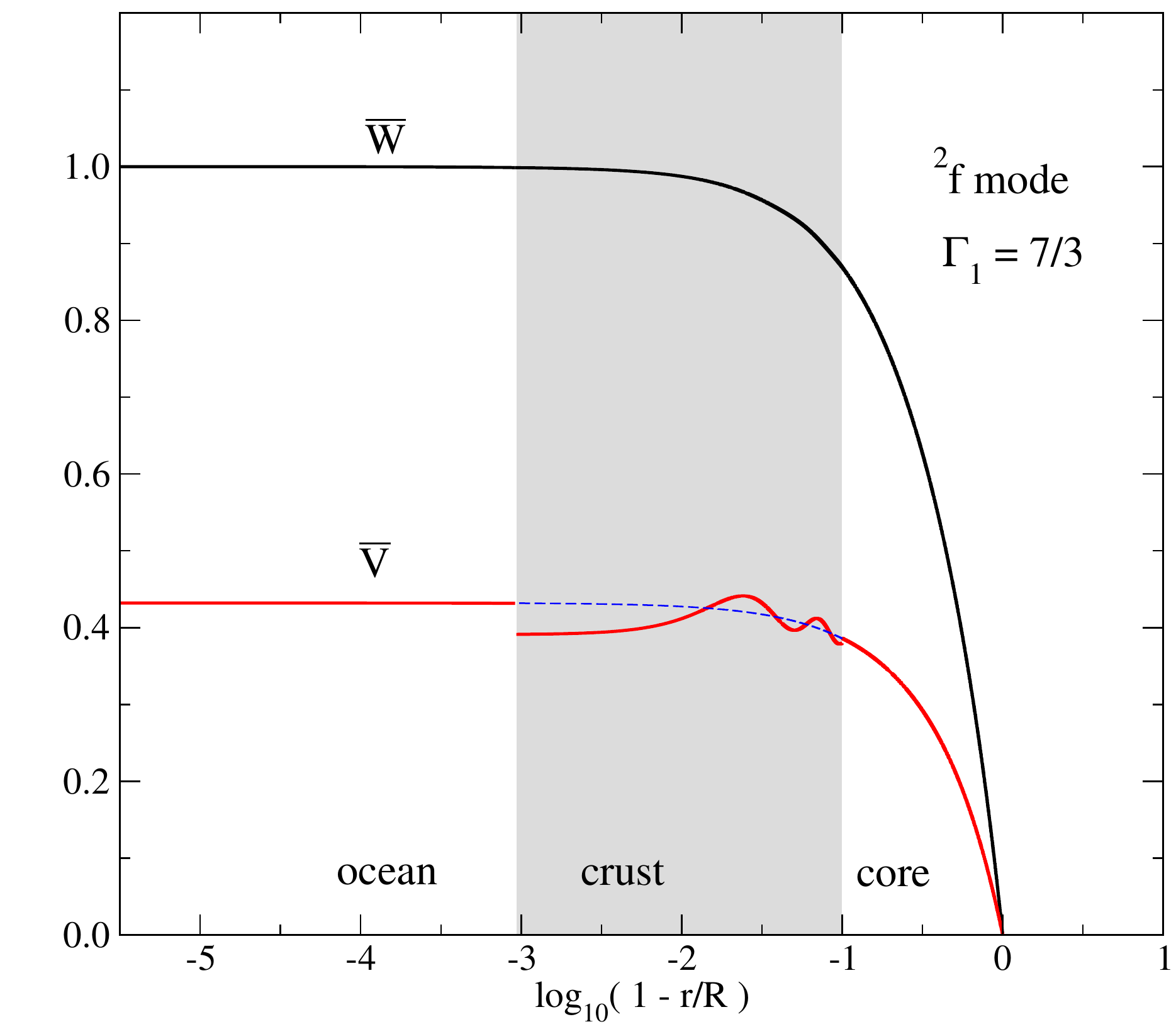} 
\caption{ \label{fig:f-mode}
Lagrangian displacements for the $l=2$ f mode. The black and red curves show, respectively, the functions  $\overline{W} = W / (r R )$ and  $\overline{V} = V / (r R )$  for a model with $\Gamma_1=7/3$ and an elastic crust. The dashed blue line represents the eigenfunction for the corresponding fluid model. The results demonstrate that the crust elasticity has a small, but distinguishable, effect on the mode eigenfunctions.}
\end{center}
\end{figure}
%------------------------------------------------------------------------------%

Since we are interested in the tidal excitation of the different modes by a binary companion and the possible impact on the gravitational-wave signal, we focus our attention on the quadrupole  ($l=2$) modes. We then have the sets of  oscillation frequencies provided  (in dimensionless units)  in Tables~\ref{tab:tab1}, \ref{tab:tab2} and \ref{tab:tab3}. The tabulated data include other 
  relevant quantities required for the tidal problem (e.g. the overlap integrals, the ratio at the surface between the tangential and radial components of the Lagrangian displacement  
  and the mode contribution to the Love number, all discussed in Section~\ref{expand}). 
  The p$_n$ modes (with $n$ labelling each mode) cover the high frequency range of the spectrum---above the frequency of the f mode---and their frequencies  increase for higher 
  order  ($n$) modes. In contrast, the  (gravity) g modes cover the low frequency range of the oscillation spectrum---below the f mode---and their frequencies decrease  
  for higher order modes. As a result, the g-mode spectrum becomes progressively dense towards zero frequency.  This makes the identification of the individual high-order g modes challenging.
  A stratified ocean  also sustains a family of g modes, here referred to as surface g modes (g$^s$). These also  have 
  low frequencies and their eigenfunctions are mainly confined to the star's ocean. 
  The shear mode frequency scales approximately as $\omega_\n \sim \sqrt{\check\mu / \rho }$ and increases for higher order modes. 
   Finally, the interface modes are characterised by  their low frequency and a distinctive cusp in the radial displacement at the relevant interface. 
  
The main contribution to  the dynamical tide is provided by the f mode \citep{hind1,stein,ap20a, ap20b}. This mode is known to be 
weakly affected by composition stratification \citep[see the results of][]{ap20a} and crust elasticity \citep{mcd88}. 
This is demonstrated by the results in Figure~\ref{fig:f-mode}, where we show the $l=2$ f-mode eigenfunction for a strongly stratified model with $\Gamma_1=7/3$. 
The radial component ($W$) of the Lagrangian displacement is practically equal to the barotropic case, while the 
 tangential component ($V$) exhibits  oscillations in the crust region (compared to the result for a fluid star).
 Note that, in stars with a crust the tangential displacement may be discontinuous at the fluid-elastic transitions.  
 This is also evident from Figure~\ref{fig:f-mode}.

%--------------------FIG. 2-----------------------------%
\begin{figure}
\begin{center}
\includegraphics[height=70mm]{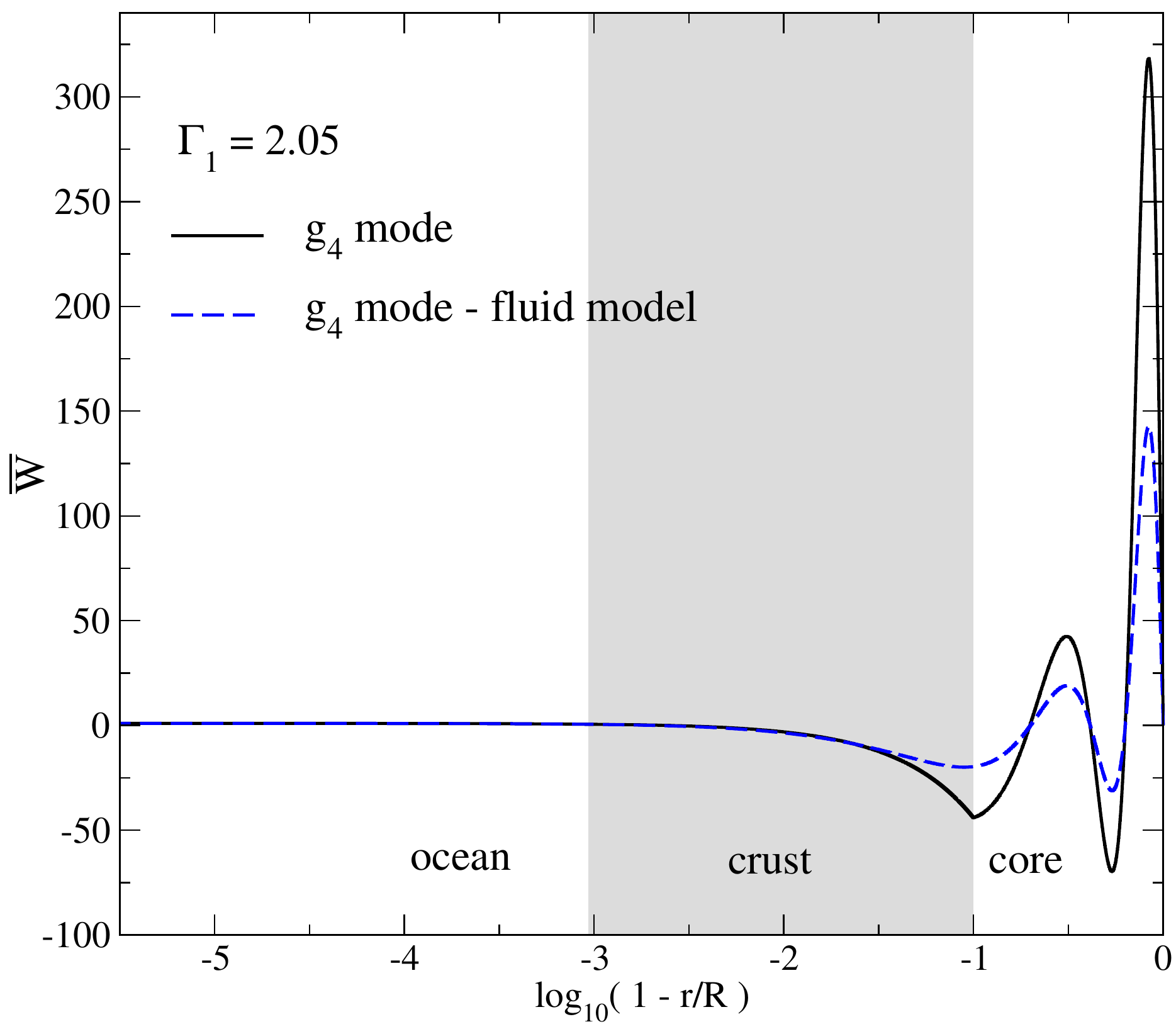} 
\includegraphics[height=70mm]{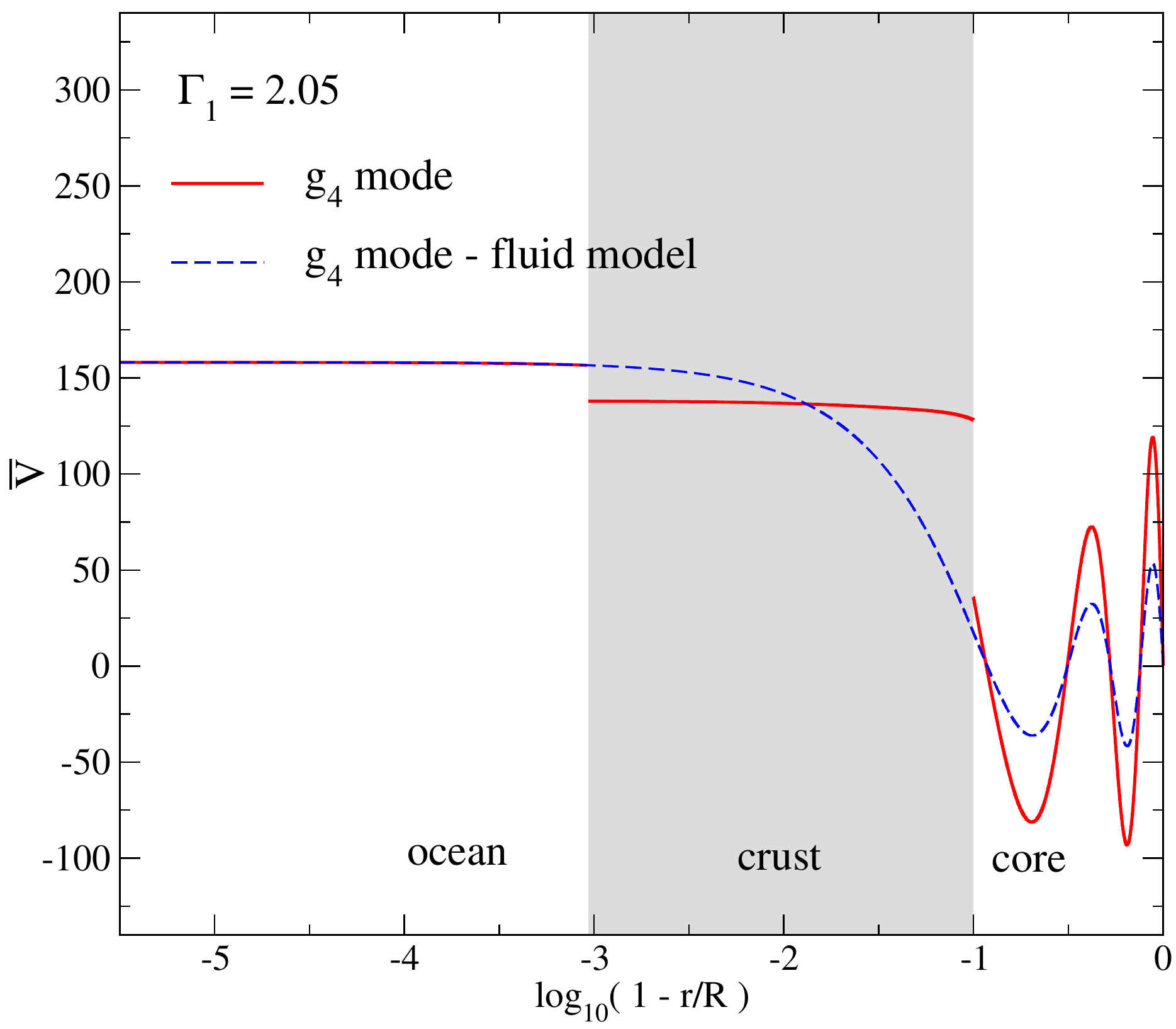} 
\caption{Lagrangian displacements for the g$_4$ mode. The two panels show, respectively,  $\overline{W} = W / (r R )$ (left panel) and 
 $\overline{V} = V / (r R )$  (right panel). The solid lines indicate the solutions for the $\Gamma_1=2.05$ star with a crust while the dashed blue line represents the corresponding fluid case. 
 \label{fig:g4-mode}}
\end{center}
\end{figure}
%------------------------------------------------%

%-----------------FIG. 3--------------------------%
\begin{figure*}
\begin{center}
\includegraphics[height=51mm]{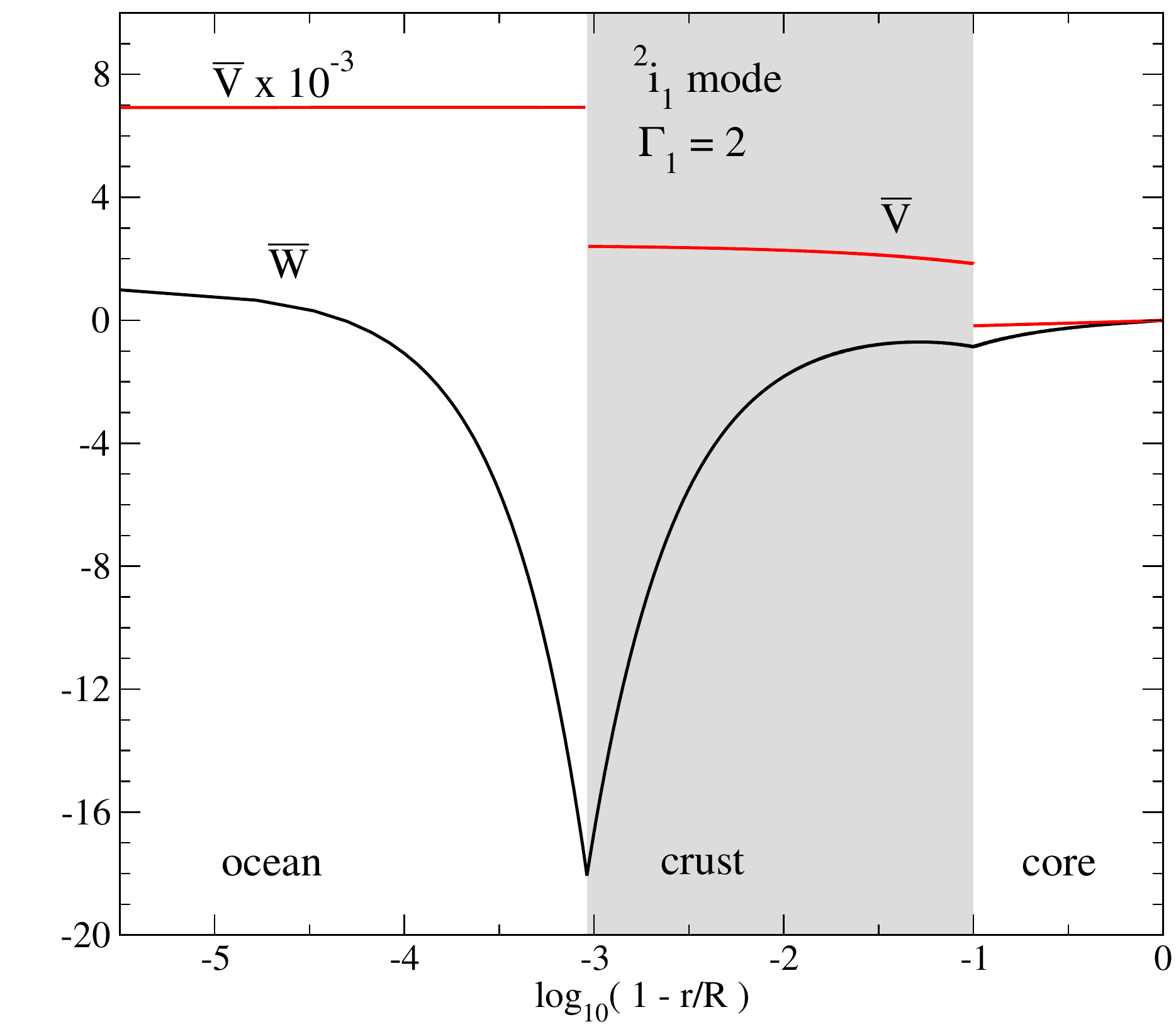} 
\includegraphics[height=51mm]{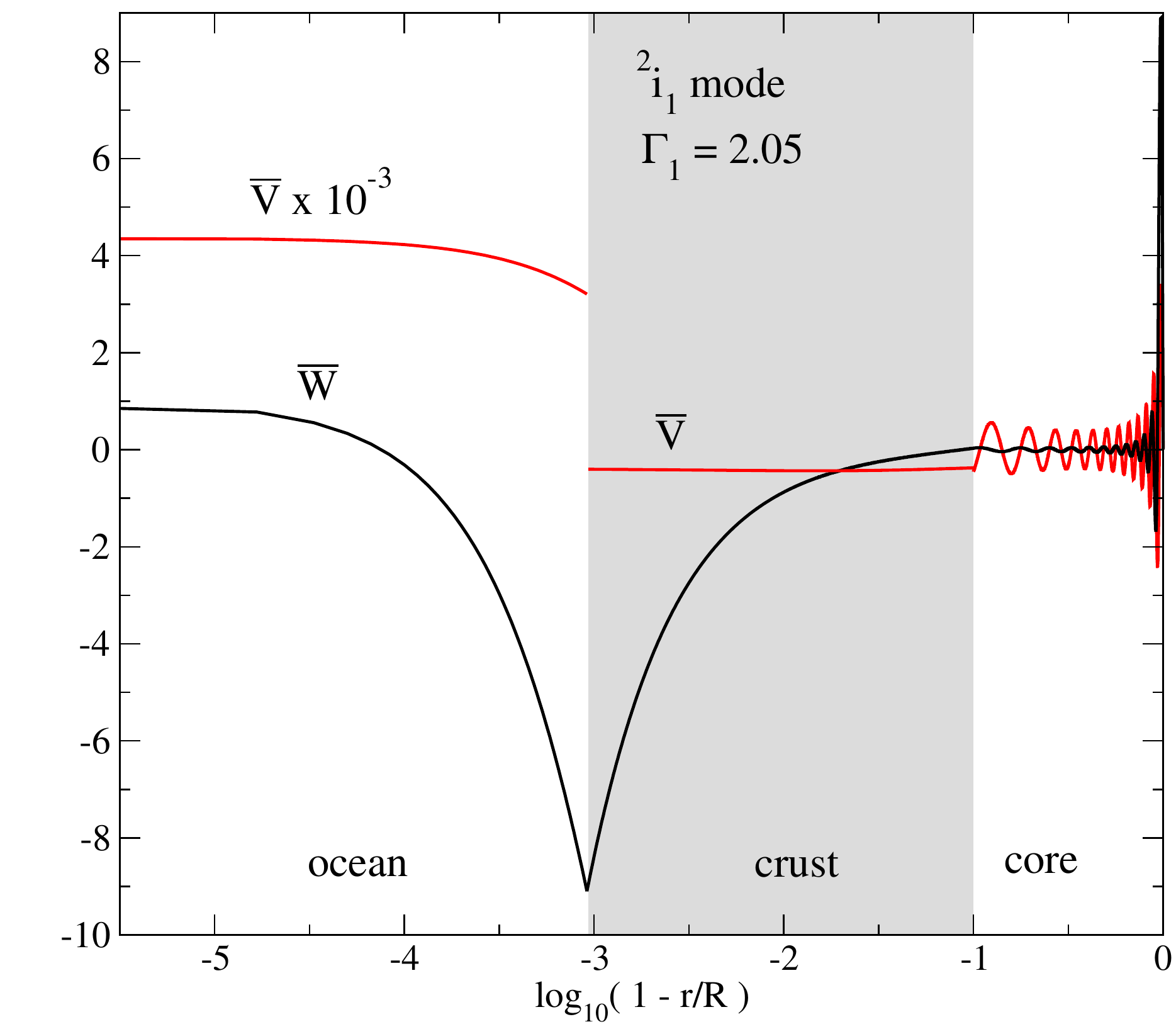} 
\includegraphics[height=51mm]{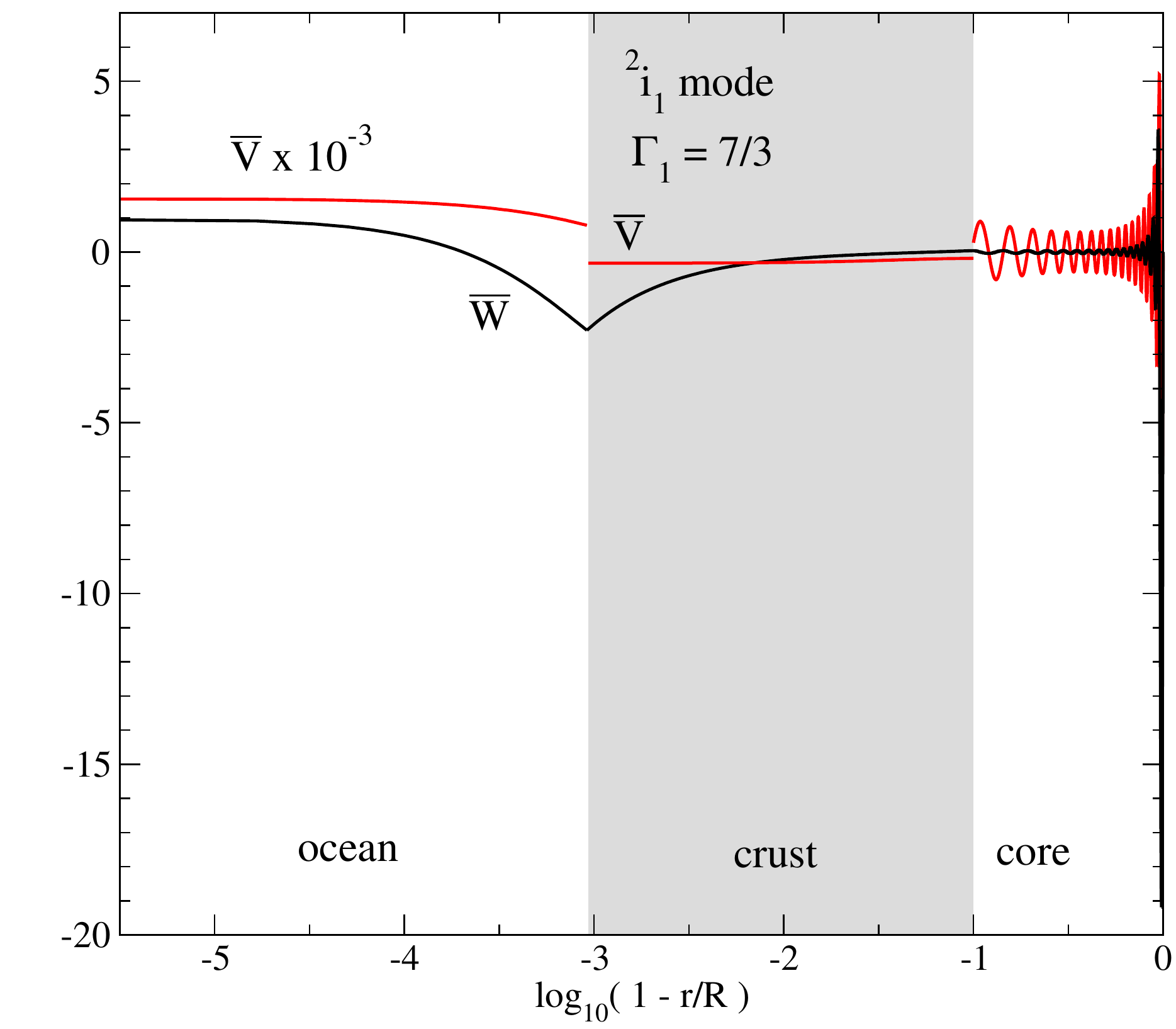}
\caption{Eigenfunctions for the i$_1$ mode, which is associated with the crust-ocean transition (note the distinctive cusp in the radial displacement). The black curves represent the radial component $\overline{W} = W / (r R )$  while the red curves are for the tangential component $\overline{V} = V / (r R )$. The left-hand panel shows the results for a barotropic model ($\Gamma_1=2$). The middle  panel represents the $\Gamma_1=2.05$ case, while the right-hand panel is for a stratified model with $\Gamma_1=7/3$.
    \label{fig:iA}}
\end{center}
\end{figure*}
%----------------------------------------------------%

%--------------------FIG. 4-----------------%
\begin{figure*}
\begin{center}
\includegraphics[height=70mm]{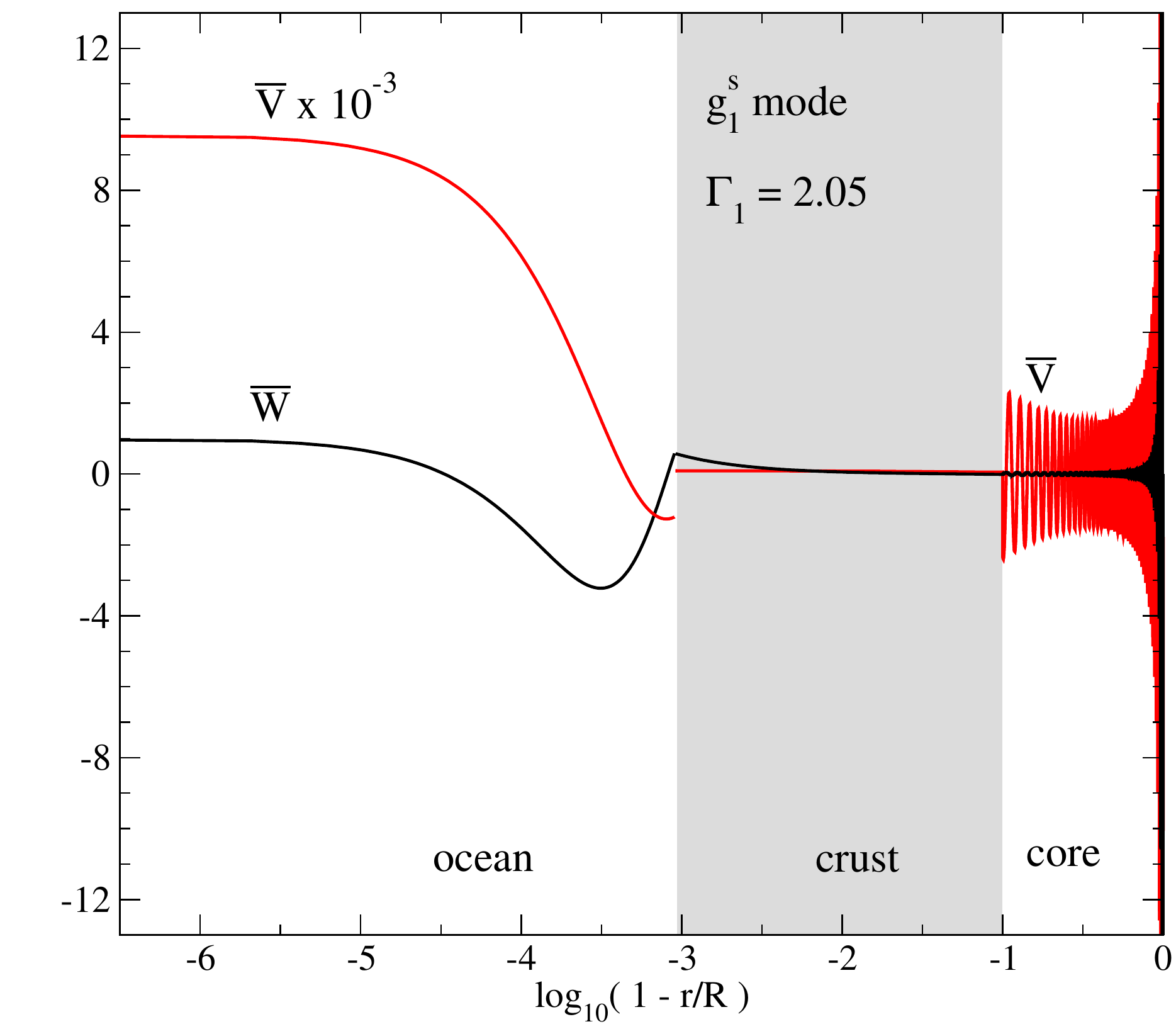} 
\includegraphics[height=70mm]{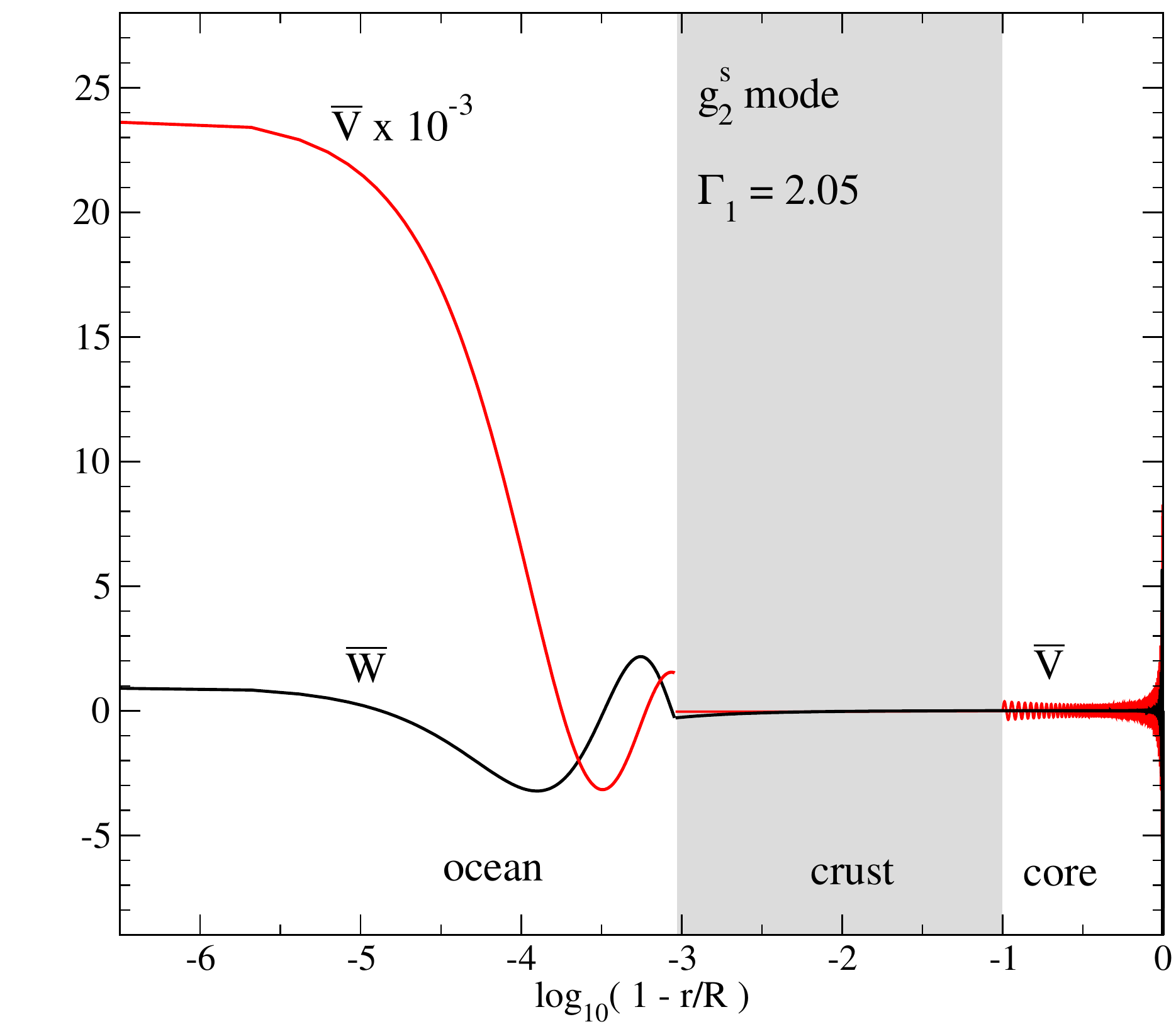} 
\caption{Eigenfunctions for the first two surface g modes for the stratified model with $\Gamma_1=7/3$. 
The left-hand panel shows the $\overline{W}$ and $\overline{V}$ eigenfunctions for the g$^{\s}_1$ mode, while 
right-hand panel displays the same quantities for the  g$^{\s}_2$ mode.   \label{fig:gs}}
\end{center}
\end{figure*}
%------------------------------------------------------%

Qualitatively similar conclusions apply to the g and p modes. As an example, we consider the  g$_4$ mode for the $\Gamma_1=2.05$ model.
In Figure \ref{fig:g4-mode} we show the Lagrangian displacements for this mode.  As before, the crust mainly impacts on  the tangential component ($V$) which appears to be almost constant in the crust.

Associated with each fluid-elastic interface we find a single interface mode. 
For these modes, the radial displacement is 
peaked at the  interface. 
This feature is evident from Figure
 \ref{fig:iA}, where we show the displacements for  the crust-ocean interface mode (i$_1$)  
for the barotropic model as well as for the weakly and strongly stratified stars.  
As anticipated, the $W$ eigenfunction has a notable cusp at the transition between the crust and the ocean. 
For the barotropic case, there is also another cusp (albeit with smaller amplitude) at the crust-core interface. As in the case of other modes,  
the tangential  eigenfunction $V$ is discontinuous at the interfaces, and reaches a very large 
amplitude at the star's surface. Note that the function $V$ has been reduced, in Figure \ref{fig:iA}, by a factor $10^{-3}$ in the ocean (only). Similar large amplitudes in the ocean were noted by \citet{mcd88}.

The stratification associated with the matter composition affects  the i-mode eigenfunctions, especially in the core, see Figure \ref{fig:iA}. 
In this region, the character of the interface modes is very
similar to that of the (core) g modes. This behaviour usually occurs when two modes lie in the same frequency range and the resulting  oscillation mode exhibits a mixed character.  
In this case, we find  g-mode features in the core and the characteristic interface mode cusp at the transition density.

 %------------------------------FIG. 5------------------------------------------%
\begin{figure*}
\begin{center}
\includegraphics[height=70mm]{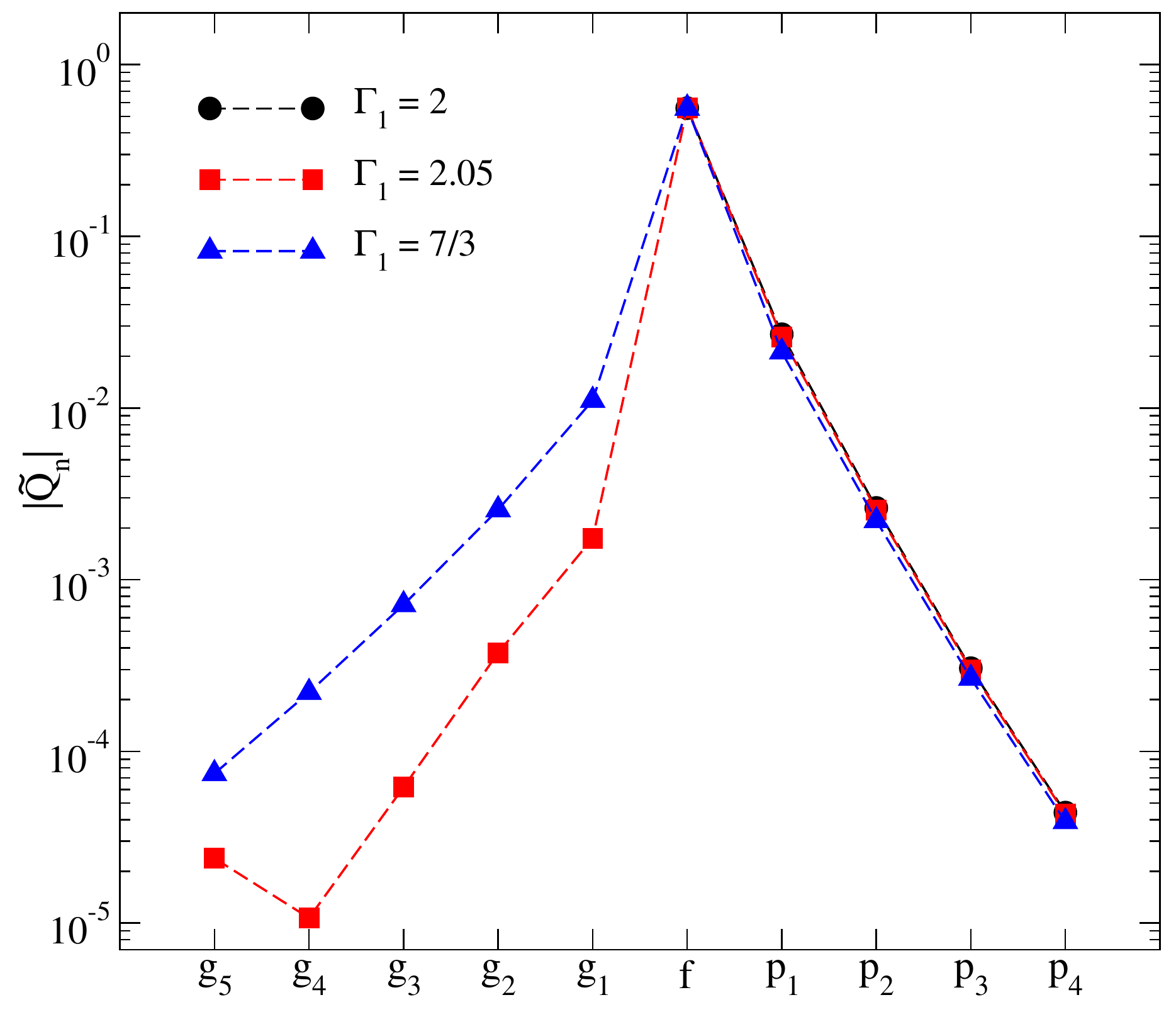} 
\includegraphics[height=70mm]{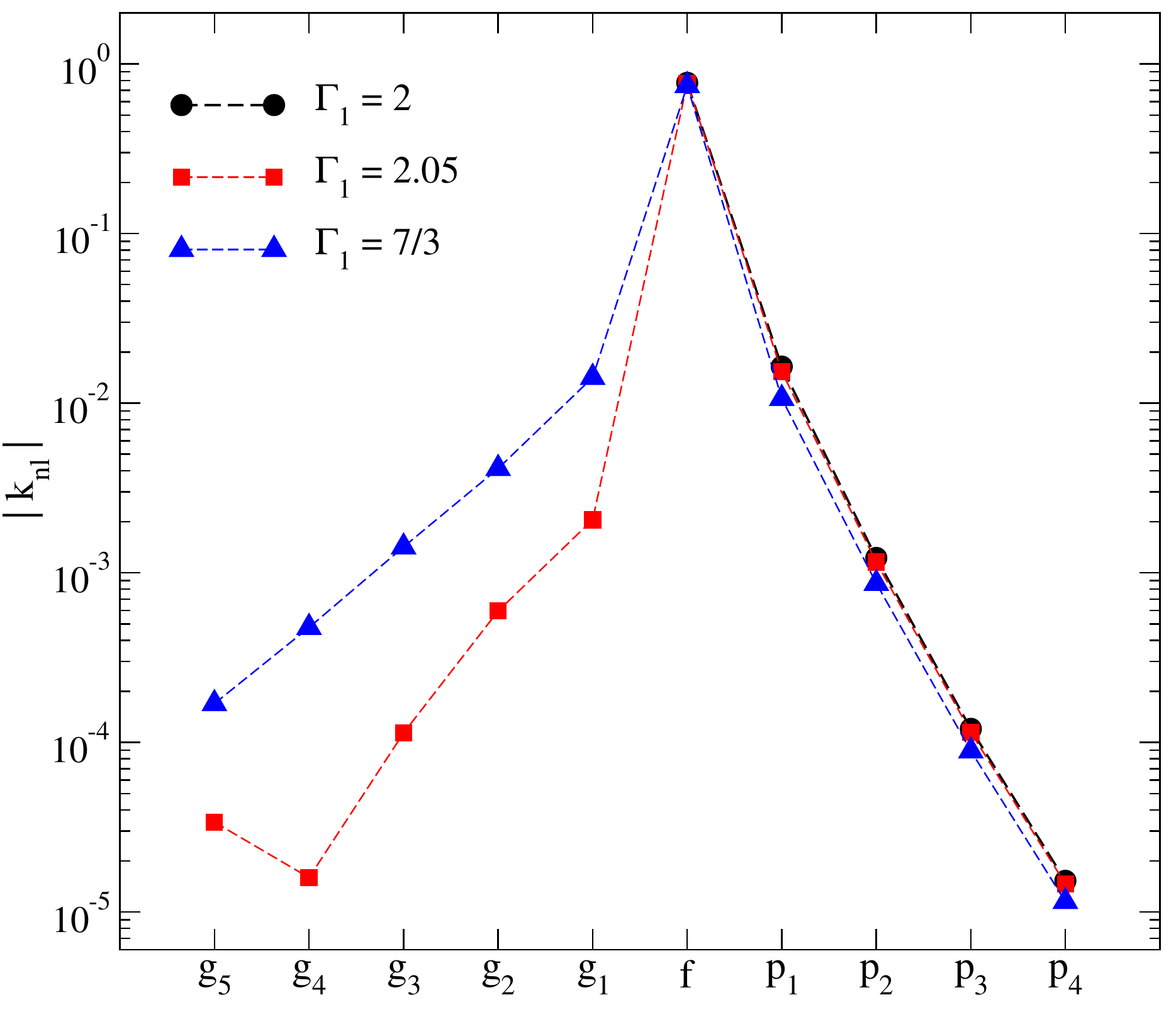} 
\caption{Overlap integrals (left panel) and mode contributions to the Love number $k_{l}^n$ (right panel) for the fundamental, pressure and gravity modes for the three stellar models we consider. 
Different markers indicate the different values of  $\Gamma_1$ (see the legend). \label{fig:Qn}}
\end{center}
\end{figure*}
%------------------------------------------------------------------------------%

The other interface mode (i$_2$), associated with the crust-core transition, is readily determined for the barotropic case ($\Gamma_1 = 2$), for which it has the expected properties. 
However, we find it difficult to identify this mode for the  stratified models. The problem is most likely due to the fact that, in the core, the i and (core) g modes have  similar eigenfunctions
 and it is difficult to numerically separate them.  This behaviour is shown in  Figure \ref{fig:g4-mode} for the g$_4$ mode, where it is clear that
the $W$ eigenfunction is not differentiable  at the crust-core transition. For higher-order g modes this behaviour is even more apparent.

The mixing of modes belonging to different classes is also notable for the surface g modes. This is illustrated in  Figure \ref{fig:gs}, where we show, for the stratified model with $\Gamma_1 = 7/3$, the $W$ and $V$ eigenfunctions for the first two g$^s$ modes.  We see that, 
in the ocean we have  the characteristic eigenfunction of a surface g mode, while the eigenfunctions $W$ and $V$ show features associated with  
a  high-order g mode in the core. The crust region appears to ``isolate'' the two regions which makes it difficult to numerically establish which set of modes a given solution belongs to. It is not even clear that this is a meaningful question for the high overtone modes.

%%%%%%%%%%%%
\subsection{The Love number}

The tidal response of a neutron star is closely related to the nature of the  different oscillation modes. This is natural since the modes form a complete set and, hence, can be used as a basis to express the  behaviour of the stellar fluid. In the static limit the sum over the star's oscillation modes leads to 
the Love number, as explicitly demonstrated by  \citet{ap20a}. The mode-sum also provides a handle on the dynamical tide (through the effective, frequency dependent, Love number $k_l^\mathrm{eff}$ from Section~\ref{expand}). In particular, 
during a binary inspiral some of the oscillation modes may pass through  resonance with the tidal driving and as a result
reach a significant amplitude.

%------------------------------FIG. 6------------------------------------------%
\begin{figure*}
\begin{center}
\includegraphics[height=70mm]{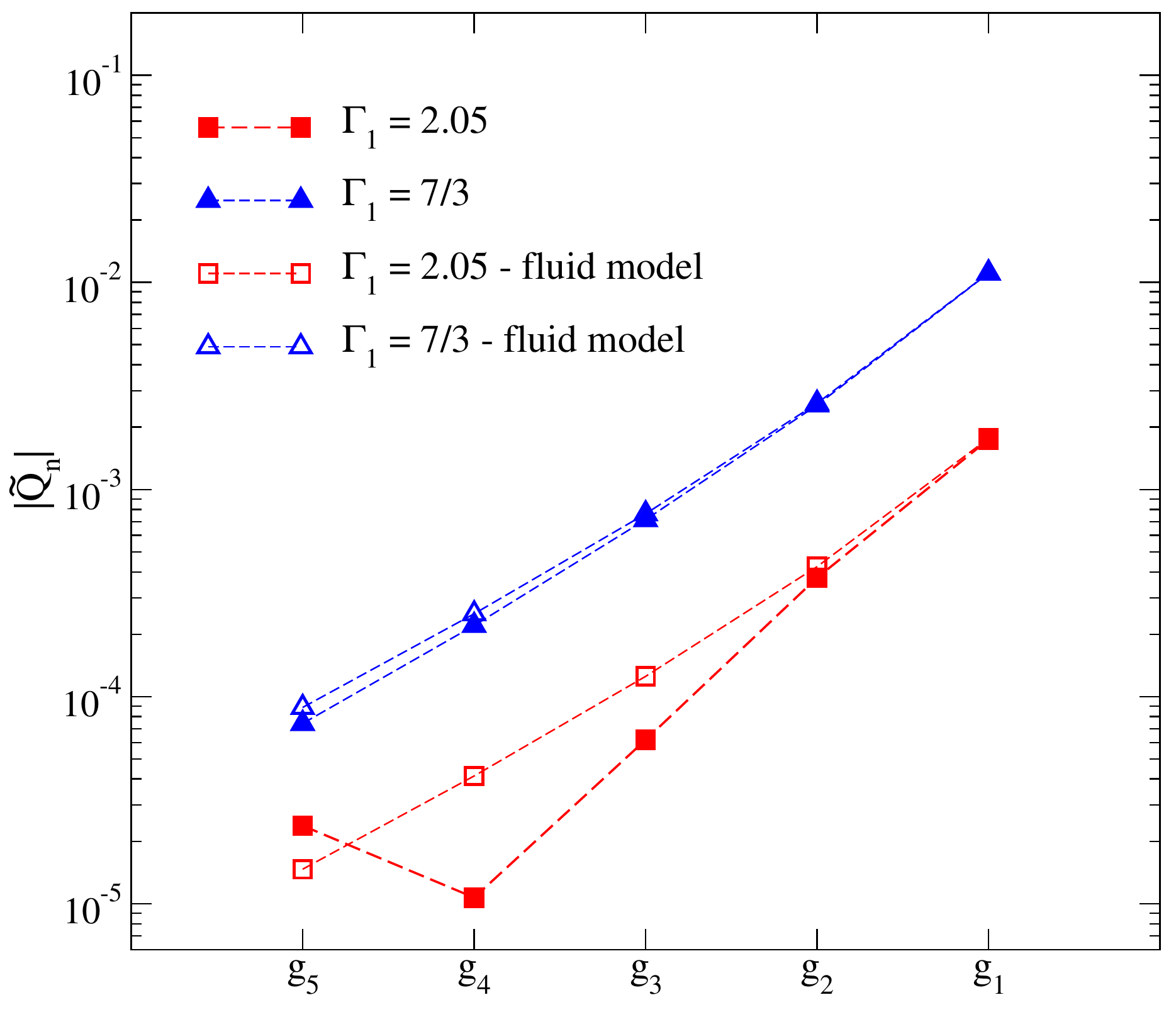}
\includegraphics[height=70mm]{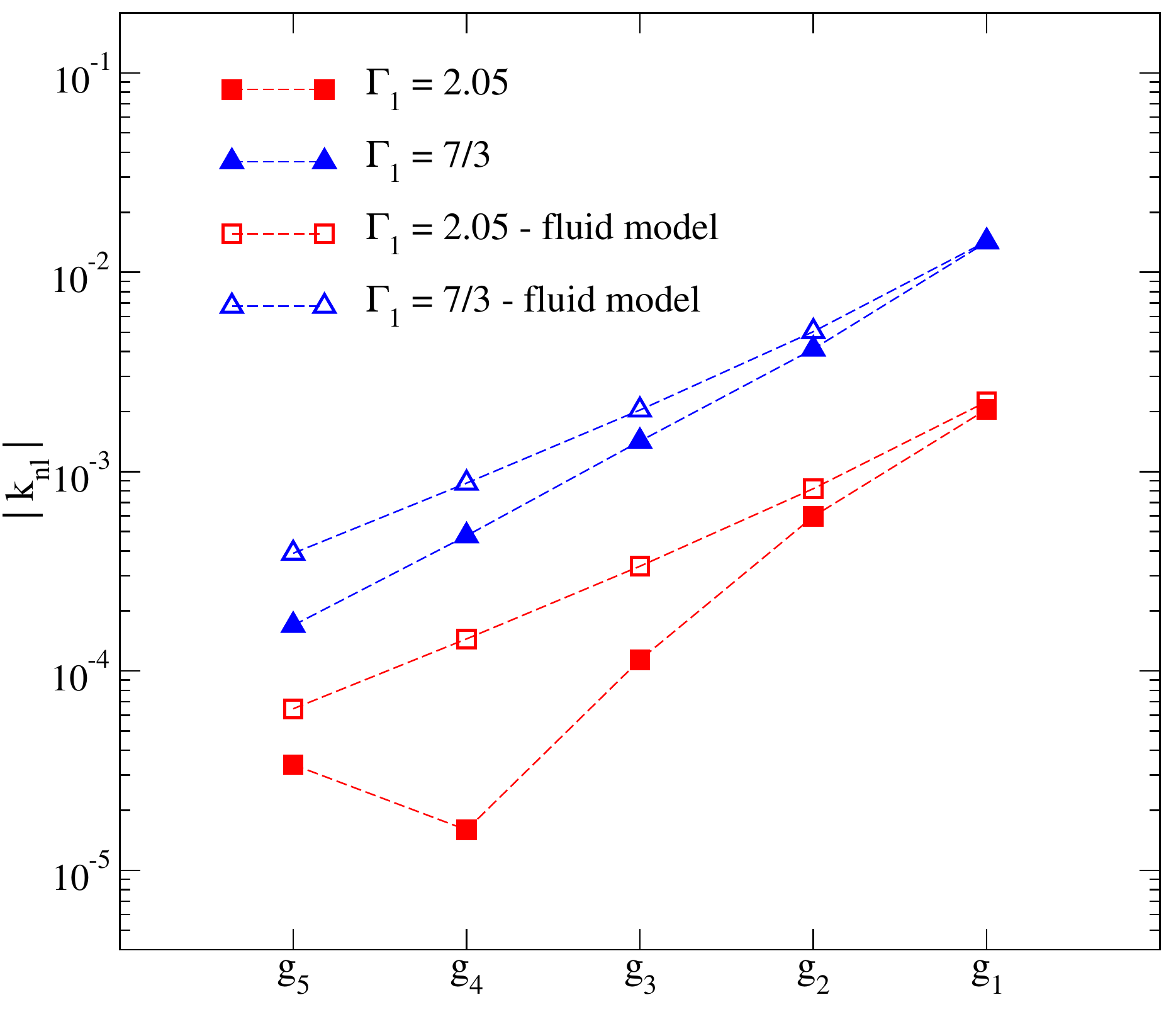} 
\caption{Comparing  models with and without a crust. We show the (dimensionless) overlap integral $Q_n$ (left panel) and the mode contribution to the Love number $k_{l}^n$ (right panel) for the first five gravity modes. 
The results for purely fluid stars are shown as  empty markers, while the results for models with crust and ocean are shown as filled markers (see the legend). As a general trend, each quantity decreases for the higher order modes, but there are clearly exceptions to this. \label{fig:cmp-Qn-knl}}
\end{center}
\end{figure*}
%------------------------------------------------------------------------------% 
%------------------------------FIG. 7-----------------------%
\begin{figure}
\begin{center}
\includegraphics[height=70mm]{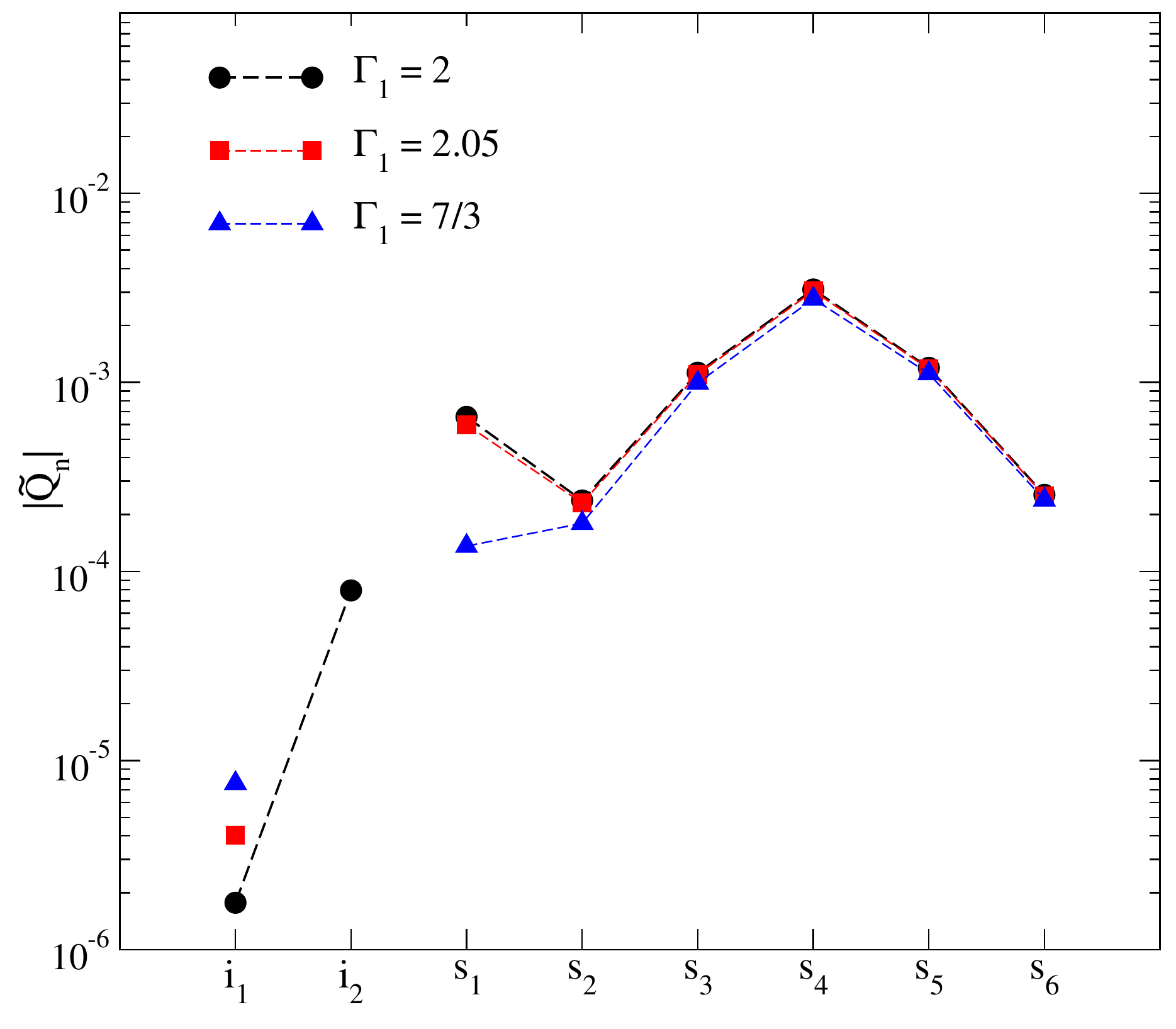} 
\includegraphics[height=70mm]{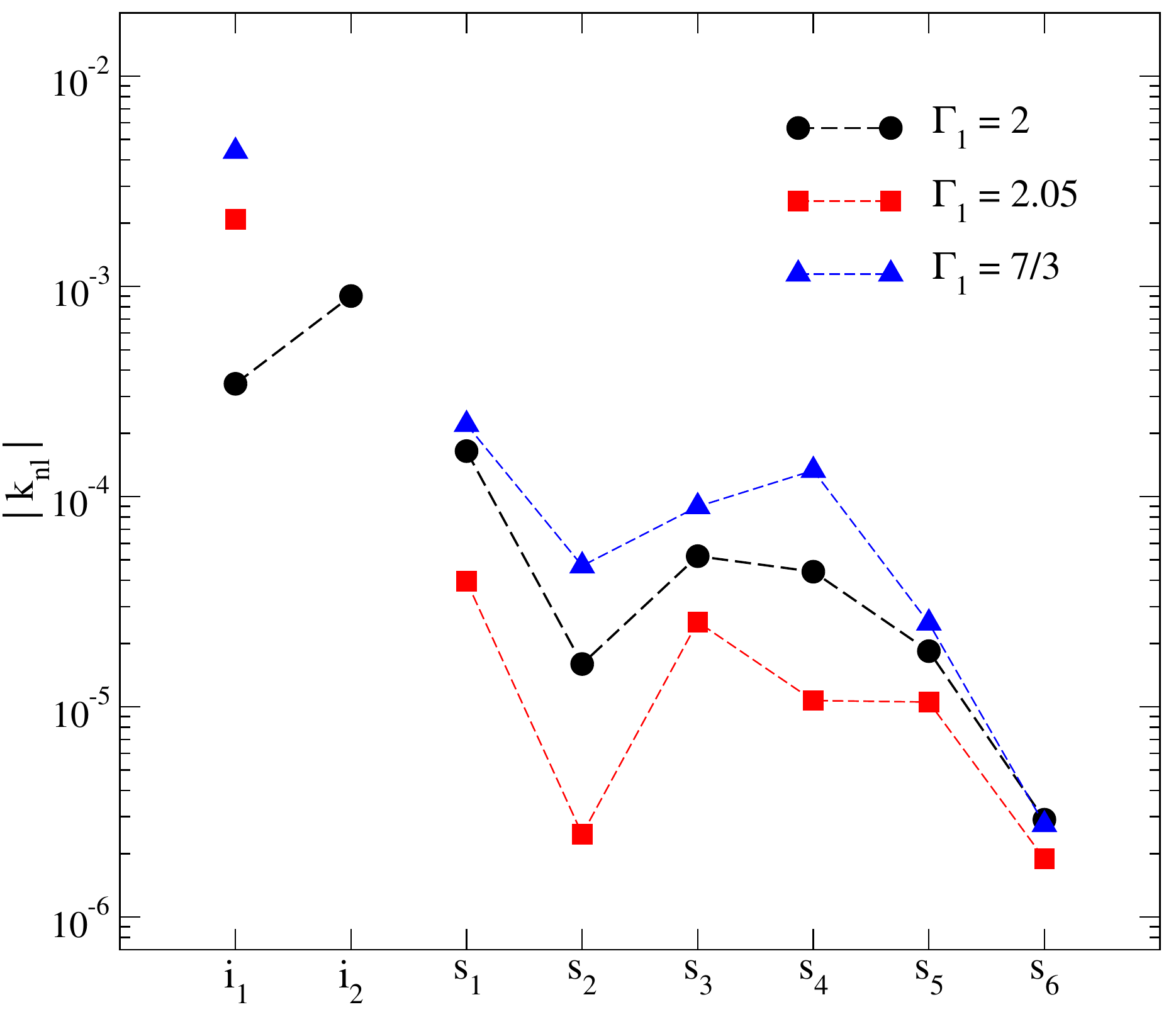} 
\caption{Overlap integrals (left panel) and mode contributions to the Love number $k_{l}^n$ (right panel) for the shear and interface modes for the three stellar models we consider. 
Different markers indicate the different values of  $\Gamma_1$ (see the legend).
  \label{fig:knl}}
\end{center} 
\end{figure}
%-------------------------------------------------------%

Let us first consider the static limit. In this case  oscillation modes are (clearly) not in resonance with the orbital motion. Nevertheless, we can use the mode-sum to represent the tidal response. This may seem a somewhat odd way to go about it, given that the result we want is contained in the usual Love number $k_l$ which can be calculated in a much simpler way \citep{hind0}. However, the mode representation provides valuable additional insight. In particular, it brings out the expectation that the main contribution to the tidal response is provided by the f mode, which has the best  overlap with the tidal driving force. This also leads to the question of the level at which other modes, which may depend on the matter composition etcetera, contribute. For example, we know that the elastic crust sustains shear modes. These are expected to have small overlap integrals and therefore have little impact. However, even if these expectations are true, it is useful to quantify what we mean by ``small'' and to what extent we can safely neglect the contribution of these modes. We need to be mindful of the fact that, even if the contribution from each mode is too small to impact on the gravitational-wave signal, the presence of an excited mode may have other repercussions. For example, it is possible that a mode passing through resonance reaches an amplitude where it  leads to fracturing of the crust \citep{tsang,bonga}. We will return to this possibility later.
Finally, we already know from the eigenfunctions we have provided that the  presence of  the crust affects the properties of all modes (see  Section \ref{sec:modes}). This means that there will be a (probably weak, but nevertheless) impact on the respective overlap integral and the contribution to the tidal response.

\begin{figure*}
\begin{center}
\includegraphics[height=70mm]{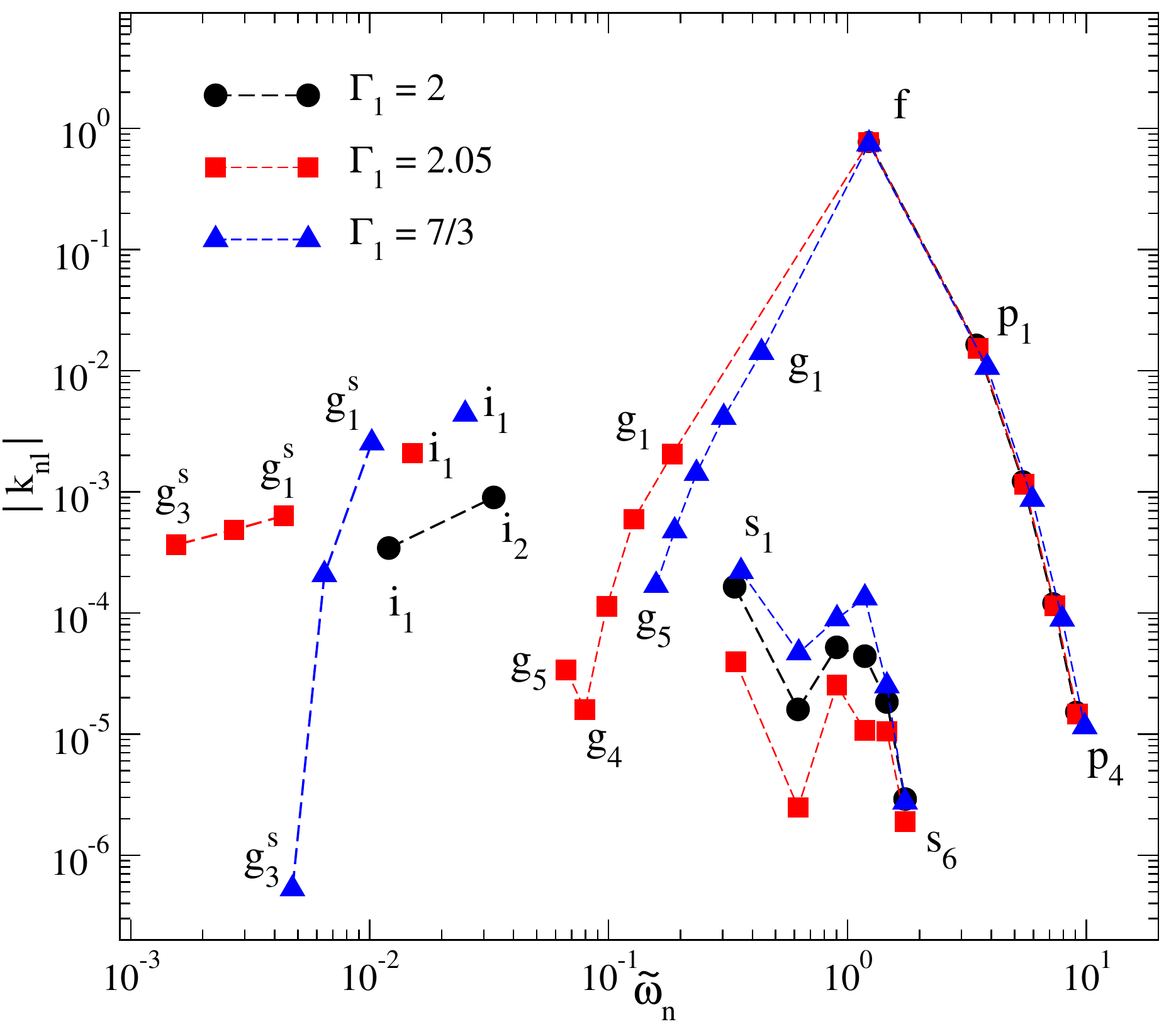} 
\caption{Summary results for all the modes we have considered. We show the Love number contribution, $k_{l}^n$,  against the mode frequencies (in dimensionless units). The illustrated modes are the fundamental (f) mode, the pressure (p$_n$) and gravity (g$_n$) modes, the shear (s$_n$) modes and finally the interface (i) modes which arise from the core-crust and crust-ocean interfaces. \label{fig:knl-omega}}
\end{center} 
\end{figure*}

As explained in Section~\ref{expand}, in the low-frequency limit (low in the sense that $\omega \ll \omega_n$), 
the tidal Love number may be written as a sum over individual mode contributions $k_{l}^{n}$ (see Equation \eqref{klfinal}). Hence, we report in Tables \ref{tab:tab1}, \ref{tab:tab2} and \ref{tab:tab3}  the value of $k_{l}^{n}$ for the different oscillation modes (along with other relevant quantities). The results confirm that the f mode dominates the tidal response. It has the largest value for the overlap integral and makes the dominant contribution to the Love number.
The contributions from the pressure and gravity modes become less important for the higher overtones (increasing  $n$), in accordance with the results of \citet{ap20a} (although there are exceptions to this, see Figures \ref{fig:Qn} and \ref{fig:knl-omega}).
For strongly stratified models, with $\Gamma_1 = 7/3$, the first g mode has a value for $k_l^n$ similar to that of the first p mode.

%----------------------FIG. 8-------------------%
%----------------------------------------------------------%

Comparing the results for stellar models with and without a crust, we find that the pressure modes are essentially not affected at all, while the
gravity modes have a less regular behaviour.  As shown in Figure~\ref{fig:cmp-Qn-knl}, 
the contribution of the g modes to the  Love number, $k_l^n$, tends to be smaller for the model with a crust (although it is  practically unchanged for the g$_1$ mode). 
In strongly stratified models, the g modes are  less influenced by the crust, and the corresponding values of $Q_n$  and $k_l^n$ are similar to the fluid case, most likely because the buoyancy dominates the elastic restoring force.

The shear modes in the crust are, as expected, more or less irrelevant for the tidal problem, as confirmed by the small  values for $Q_n$ and $k_l^n$. In Figure \ref{fig:knl} we show the two quantities for the first six s modes, where we see that the overlap integrals have similar values in the barotropic and the stratified models, while the behaviour for $k_l^n$ is less regular; the values for the $\Gamma_1 = 2.05$ model are smaller than the ones for the barotropic and strongly stratified models.
Meanwhile, the interface modes linked to the core-crust and crust-ocean transitions may contribute to the tidal response at a level similar to that of the first or second g modes (see Figures \ref{fig:knl} and \ref{fig:knl-omega}). The importance of these modes increases for models with stronger stratification, although
  their overlap integrals remain very small (see the left panel of Figure \ref{fig:knl}).

An overall summary view of the results for the Love number is provided in Figure \ref{fig:knl-omega}, where we show the quantity $k_l^n$ for all  modes  considered in this work.  
From this figure we infer which are the most relevant modes in the static limit and in which order the modes will be excited during a  
binary inspiral (as the driving frequency increases). In this figure we also show $k_l^n$ for the first three surface g modes, which mainly reside in the ocean. 

Finally, from a technical perspective, it is worth noting that some of the  oscillation modes have very small overlap integrals, the  calculation of which may be subject to numerical errors. 
This is particularly the case for high-order modes which tend to have many nodes in their eigenfunctions, leading to cancellations in the calculation of the overlap integral. 
To monitor the numerical errors we determine $Q_n$ from equations  (\ref{overlap}),  (\ref{multip}) as well as the (equivalent) expression %\citep{wein}
\citep{l94}
\begin{equation}
Q_n = l \int \rho \left[ W + \left( l + 1 \right) V \right] r^l dr \, . \label{eq:Qn_ter}
\end{equation}
As we have already mentioned, in order to increase the precision of the calculation,  we solve the perturbation equations first with a multiple shooting method and then with a relaxation step. 
As expected, the solutions obtained from the relaxation are more accurate, which allows us to extract the high-order  modes. A few additional comments on the technical aspects of the calculation are provided in  Appendix \ref{sec:num}.

\begin{table*}
\begin{center}
\caption{Mode results for the barotropic model with $\Gamma_1 = 2$ \label{tab:tab1} }
\begin{tabular}{c c c c c  }
\hline
 $ \rm Mode $ &  $ \tilde \omega_n $  &  $ |Q_n| $   & $ \left( V/W \right)_R $  & $ k_{nl}$ \\ 
\hline
 p$_4$    &  9.0525       &	 $4.3939 \times 10^{-5}$  &  $1.2202\times 10^{-2}$   &   $ \hspace{0.3cm}1.5267 \times 10^{-5} $ \\
 p$_3$    &  7.2615  	    &	 $3.0496 \times 10^{-4}$  &  $1.8965\times 10^{-2}$   &   $-1.2023  \times 10^{-4} $ \\
 p$_2$    &  5.4158       &   $2.6171\times 10^{-3}$   &  $3.4085\times 10^{-2}$   &   $ \hspace{0.3cm}1.2234  \times 10^{-3} $ \\
 p$_1$    &  3.4618       &	  $2.6879\times 10^{-2}$  &  $8.3830\times 10^{-2}$   &   $-1.6431  \times 10^{-2} $ \\ 
 f             &  1.2269       &    $5.5791\times 10^{-1}$  &  $4.4142\times 10^{-1}$   &   $  \hspace{0.3cm}7.7453 \times 10^{-1} $ \\ 
 s$_1$    &  0.3368       &	  $6.5765\times 10^{-4}$   &  $8.5595                      $   &   $ \hspace{0.3cm} 1.6464 \times 10^{-4} $ \\
 s$_2$    &  0.6204       &	  $2.3746\times 10^{-4}$   &  $2.6286                      $   &   $ -1.5983 \times 10^{-5} $ \\
 s$_3$    &  0.9014       &	  $1.1257\times 10^{-3}$   &  $1.1843                      $   &   $ \hspace{0.3cm} 5.2074 \times 10^{-5} $ \\ 
 s$_4$    &  1.1815       &	  $3.1028\times 10^{-3}$   &  $0.8578                      $   &   $ -4.3958 \times 10^{-5} $ \\  
 s$_5$    &  1.4614       &	  $1.1936\times 10^{-3}$   &  $0.4895                      $   &   $ -1.8403 \times 10^{-5} $ \\  
 s$_6$    &  1.7412       &	  $2.5426\times 10^{-4}$   &  $0.3269                      $   &   $  \hspace{0.3cm}2.9057  \times 10^{-6} $ \\    

 i$_2$    &  0.0331       &	  $7.9418\times 10^{-5}$   &  $9.0802\times 10^{2}$    &   $ \hspace{0.3cm} 8.9995  \times 10^{-4} $ \\    
 i$_1$    &  0.0120       &	  $1.7696\times 10^{-6}$   &  $6.9343\times 10^{3}$    &   $ -3.4406  \times 10^{-4} $ \\
 \hline 
$k_l$ & & & &  0.25991  \\
 \hline   
\end{tabular}
\end{center}
\end{table*}
%------------------------------------------------------------%

%------------------------------TAB. 2------------------------------------------%
\begin{table*}
\begin{center}
\caption{Same as Table \ref{tab:tab1}, for the weakly stratified model with $\Gamma_1 = 2.05$.\label{tab:tab2} }
\begin{tabular}{c c c c c  }
\hline
 $ \rm Mode $ &  $ \tilde \omega_n $  &  $ |\tilde  Q_n| $   & $ \left( V/W \right)_R $  & $ k_{l}^n$ \\ 
\hline
 p$_4$    &  9.1741       &	  $4.3149 \times 10^{-5}$  &  $1.1881\times 10^{-2}$   &   $ \hspace{0.3cm}1.4615 \times 10^{-5} $ \\
 p$_3$    &  7.3626  	    &	  $2.9863 \times 10^{-4}$  &  $1.8448\times 10^{-2}$   &   $-1.1465  \times 10^{-4} $ \\
 p$_2$    &  5.4963       &    $2.5463 \times 10^{-3}$   & $3.3095\times 10^{-2}$   &   $ \hspace{0.3cm}1.1575  \times 10^{-3} $ \\
 p$_1$    &  3.5206       &	  $2.5858\times 10^{-2}$  &  $8.1036\times 10^{-2}$   &   $-1.5319 \times 10^{-2} $ \\

 f             &  1.2274       &    $5.5795\times 10^{-1}$  &  $4.3989 \times 10^{-1}$   &   $  \hspace{0.3cm}7.6983 \times 10^{-1} $ \\

 g$_1$    &  0.1848       &	  $1.7435 \times 10^{-3}$  &  $2.7686\times 10^{1}$   &   $ \hspace{0.3cm} 2.0546 \times 10^{-3} $ \\
 g$_2$    &  0.1277      &	  $3.7451\times 10^{-4}$  &  $6.0184\times 10^{1}$   &   $ \hspace{0.3cm} 5.9549 \times 10^{-4} $ \\
 g$_3$    &  0.0983       &	  $6.1782\times 10^{-5}$  &  $1.0300\times 10^{2}$   &   $ \hspace{0.3cm} 1.1358 \times 10^{-4} $ \\ 
 %g$_4$    &  0.0796       &	  $7.2087\times 10^{-4}$  &  $1.5811\times 10^{2}$   &   $ -1.5940 \times 10^{-5} $ \\  
 g$_4$    &  0.0796       &	  $1.0704\times 10^{-5}$  &  $1.5811\times 10^{2}$   &   $ -1.5940 \times 10^{-5} $ \\   
 g$_5$    &  0.0664       &	  $2.3809\times 10^{-5}$  &  $2.2786\times 10^{2}$   &   $ -3.3875 \times 10^{-5} $ \\  
 
 s$_1$    &  0.3403       &	  $5.9732\times 10^{-4}$  &  $9.4879                      $   &   $ -3.9607 \times 10^{-5} $ \\
 s$_2$    &  0.6212       &	  $2.2991\times 10^{-4}$  &  $2.7762                      $   &   $  -2.4739 \times 10^{-6} $ \\
 s$_3$    &  0.9016       &	  $1.1045\times 10^{-3}$  &  $1.1381                      $   &   $ \hspace{0.3cm} 2.5352 \times 10^{-5} $ \\ 
 s$_4$    &  1.1817      &	  $3.0456\times 10^{-3}$  &  $1.2753                      $   &   $ -1.0712 \times 10^{-5} $ \\  
 s$_5$    &  1.4615       &	  $1.1796\times 10^{-3}$  &  $0.5044                      $   &   $ -1.0537 \times 10^{-5} $ \\  
 s$_6$    &  1.7412       &	  $2.5158\times 10^{-4}$  &  $0.3254                      $   &   $  \hspace{0.3cm}1.8895  \times 10^{-6} $ \\    

 i$_1$    &  0.0151       &	  $4.0293\times 10^{-6}$  &  $4.3984\times 10^{3}$    &   $ \hspace{0.3cm} 2.0857  \times 10^{-3} $ \\    
  g$^{\rm s}_1$    &  $4.3665  \times 10^{-3}$     &	  $3.2388 \times 10^{-7}$  &  $ 5.2450\times 10^{4}$   &   $ - 6.3519  \times 10^{-4} $ \\
 g$^{\rm s}_2$    &  $2.7049  \times 10^{-3}$     &	  $1.2203 \times 10^{-7}$  &  $1.3667  \times 10^{5}$   &   $ \hspace{0.3cm} 4.8364  \times 10^{-4} $ \\
 g$^{\rm s}_3$    &  $1.5483  \times 10^{-3}$     &	  $4.0159 \times 10^{-8}$  &  $4.1716  \times 10^{5}$   &   $ \hspace{0.3cm} 3.6536  \times 10^{-4} $ \\ 
\hline 
$k_l$ & & & &  0.26055  \\
 \hline   
\end{tabular}
\end{center}
\end{table*}
%------------------------------------------------------------------------------%
%------------------------------TAB. 3------------------------------------------%
\begin{table*}
\begin{center}
\caption{Same as Table \ref{tab:tab1}, for the strongly stratified model with $\Gamma_1 = 7/3$. \label{tab:tab3} }
\begin{tabular}{c c c c c  }
\hline
 $ \rm Mode $ &  $ \tilde \omega_n $  &  $ |\tilde  Q_n| $   & $ \left( V/W \right)_R $  & $ k_{l}^n$ \\ 
\hline
 p$_4$    &  9.8354       &	  $3.8918 \times 10^{-5}$  &  $1.0337\times 10^{-2}$   &   $ \hspace{0.3cm}1.1519 \times 10^{-5} $ \\
 p$_3$    &  7.9117  	    &	  $2.6649 \times 10^{-4}$  &  $1.5976\times 10^{-2}$   &   $-8.9014  \times 10^{-5} $ \\
 p$_2$    &  5.9329       &    $2.2059 \times 10^{-3}$   & $2.8404\times 10^{-2}$   &   $ \hspace{0.3cm}8.6638  \times 10^{-4} $ \\
 p$_1$    &  3.8413       &	  $2.1191\times 10^{-2}$  &  $6.8014\times 10^{-2}$   &    $-1.0663 \times 10^{-2} $ \\

 f             &  1.2294       &    $5.5804\times 10^{-1}$  &  $4.3209 \times 10^{-1}$   &   $  \hspace{0.3cm}7.4625 \times 10^{-1} $ \\

 g$_1$    &  0.4363       &	  $1.1076\times 10^{-2}$  &  $4.9540                     $   &   $ \hspace{0.3cm} 1.4189 \times 10^{-2} $ \\
 g$_2$    &  0.3033      &	  $2.5557\times 10^{-3}$  &  $1.0638\times 10^{1}$   &   $ \hspace{0.3cm} 4.1211 \times 10^{-3} $ \\
 g$_3$    &  0.2333       &	  $7.1476\times 10^{-4}$  &  $1.8222\times 10^{1}$   &   $ \hspace{0.3cm} 1.4202 \times 10^{-3} $ \\ 
 g$_4$    &  0.1889       &	  $2.2039\times 10^{-4}$  &  $2.7910\times 10^{1}$   &   $ \hspace{0.3cm} 4.7619 \times 10^{-4} $ \\  
 g$_5$    &  0.1580       &	  $7.4133\times 10^{-5}$  &  $3.9966\times 10^{1}$   &   $ \hspace{0.3cm} 1.6928 \times 10^{-4} $ \\  
% g$_1$    &  0.4363       &	  $1.1076\times 10^{-2}$  &  $4.9540                     $   &   $ \hspace{0.3cm} 1.4189 \times 10^{-3} $ \\
% g$_2$    &  0.3033      &	  $2.5557\times 10^{-3}$  &  $1.0638\times 10^{1}$   &   $ \hspace{0.3cm} 4.1211 \times 10^{-4} $ \\
% g$_3$    &  0.2333       &	  $7.1476\times 10^{-4}$  &  $1.8222\times 10^{1}$   &   $ \hspace{0.3cm} 1.4202 \times 10^{-4} $ \\ 
% g$_4$    &  0.1889       &	  $2.2039\times 10^{-4}$  &  $2.7910\times 10^{1}$   &   $ \hspace{0.3cm} 4.7619 \times 10^{-5} $ \\  
% g$_5$    &  0.1580       &	  $7.4133\times 10^{-5}$  &  $3.9966\times 10^{1}$   &   $ \hspace{0.3cm} 1.6928 \times 10^{-5} $ \\  

 s$_1$    &  0.3582       &	  $1.3617\times 10^{-4}$  &  $7.8000                      $   &   $  -2.2038 \times 10^{-4} $ \\
 s$_2$    &  0.6248       &	  $1.7978\times 10^{-4}$  &  $2.5559                      $   &   $ \hspace{0.3cm}  4.6861 \times 10^{-5} $ \\
 s$_3$    &  0.9029       &	  $9.9318\times 10^{-4}$  &  $1.2474                      $   &   $ -8.9661 \times 10^{-5} $ \\ 
 s$_4$    &  1.1823       &	  $2.7675\times 10^{-3}$  &  $0.6784                      $   &   $ \hspace{0.3cm} 1.3324 \times 10^{-4} $ \\  
 s$_5$    &  1.4618       &	  $1.1135\times 10^{-3}$  &  $0.4544                      $   &   $ \hspace{0.3cm} 2.5039 \times 10^{-5} $ \\  
 s$_6$    &  1.7414       &	  $2.3936\times 10^{-4}$  &  $0.3325                      $   &   $ -2.7500 \times 10^{-6} $ \\    

 i$_1$    &  0.0251       &	  $7.5769\times 10^{-6}$  &  $1.5867\times 10^{3}$    &   $ \hspace{0.3cm} 4.3880  \times 10^{-3} $ \\    
  g$^{\rm s}_1$    &   0.0102                                 &  $1.2709 \times 10^{-6}$  &  $ 9.5889 \times 10^{3}$   &   $ -2.5364  \times 10^{-3} $ \\
 g$^{\rm s}_2$    &  $6.4610   \times 10^{-3}$     &  $5.1534 \times 10^{-7}$  &  $ 2.3954 \times 10^{4}$   &   $ \hspace{0.3cm} 2.0759  \times 10^{-3} $ \\
 g$^{\rm s}_3$    &  $4.7519  \times 10^{-3}$      &  $6.0135 \times 10^{-9}$  &  $ 4.4284 \times 10^{4}$   &   $ - 5.2761  \times 10^{-7} $ \\ 
\hline 
$k_l$ & & & &  0.26057  \\
 \hline   
\end{tabular}
\end{center}
\end{table*}

%%%%%%%%%%%%%%%
\section{Crust fracturing}  
\label{sec:frac}

As  different oscillation modes pass through resonance during a binary inspiral, their amplitude may become large enough that the motion in the crust induces (local) fracturing of the nuclear lattice. It has been suggested that the interface modes are particularly relevant in this respect \citep{tsang,bonga}. It also known that the static tide is unlikely to break the crust before the binary merger \citep{penner,gittins}. In order to consider this problem we need to complement our mode analysis with the energy deposited in each mode during inspiral. This analysis closely follows, for example, \citet{l94} so we will only outline the steps here. Some further details are provided in Appendix \ref{sec:mod-res}.

%%%%%
\subsection{The mode excitation}
\label{sec:dyn}

The binary separation, $D$, shrinks at a rate which can be described, to leading order, by  
\begin{equation}
\dot D = - \frac{64}{5} \frac{G^3}{c^5} \frac{M' M_\star (M_\star+M')}{D^3} \, ,
\end{equation}
where $M'$ is  the mass of the companion. This leading-order expression should be sufficient as long as we are not trying to resolve the fine details of the problem. We already know that the effects of the tide on the orbital evolution enter at (much) higher post-Newtonian order \citep{l94,ks,hind}. 
The orbital frequency $\Omega$ follows from Kepler's law, so we have
\begin{equation}
\Omega = \left[  \frac{G  (M_\star+M')}{D^3}  \right]^{1/2} \label{eq:orb_om} \, .
\end{equation}

We know from the perturbation analysis that the mode amplitude can be calculated from Equation \eqref{eq:eqa}, where we need the tidal potential  
\begin{equation}
\chi = - G M' \sum_{l\ge 2 } \sum_{m=-l}^{m=l} \frac{W_{lm} r^l }{D(t)^{l+1}} Y_{lm} e^{- i m \Phi(t)}
\end{equation}
with $\Phi = \int \Omega(t) \mathrm{d}t$ and explicitly defining the $v_l$ coefficient we used earlier. For $l=2$ the  $W_{lm}$  coefficients have the following values: 
\begin{equation}
W_{20} =  - \sqrt{\frac{\pi}{5}} \, , \qquad    W_{2\pm1} =  0 \, ,  \qquad W_{2\pm2} =   \sqrt{\frac{3 \pi}{10}}  \, .
\end{equation}

For a given mode $(l,m)$, Equation \eqref{eq:eqa}  leads to
\begin{equation}
\ddot a_n + \omega_n ^2 a_n =    \frac{G M' }{R^3}  \frac{\tilde Q_n }{\tilde {\mathcal A}_n^2}  \left( \frac{R }{D(t)} \right)^{l+1}   W_{lm} \, e^{- i m \Phi(t)} 
\label{eq:eq_a3}
\end{equation}
where $\tilde Q_n$  is the dimensionless overlap integral defined in equation (\ref{eq:dimQn})  
and $\tilde {\mathcal A}_n^2 = \mathcal{A}_n^2 / (M_\star R^2)$. 

Equation (\ref{eq:eq_a3}) is a forced harmonic oscillator for the mode amplitude $a_n$, where the forcing term is provided by the tidal potential. 
An oscillation mode is in resonance with the orbit when $\omega_n \simeq m \Omega$. 
For the most relevant modes,  the resonance  occurs at the late stages of  binary inspiral, when the  separation  changes rapidly
and hence the energy transfer to the mode is limited.
Using the resonance condition in equation (\ref{eq:orb_om}) we have the  (dimensionless)
resonance distance \citep{l94}
\begin{equation}
\frac{D}{R} =  \left[  \frac{m^2 (1+q)}{\tilde \omega_n^2}  \right]^{1/3}
\end{equation}
where $q=M'/M_\star$ and $\tilde \omega_n$ is the dimensionless mode frequency.

For each oscillation mode, we can determine the kinetic and potential energy from
\begin{align}
 E_k (t) & = \frac{1}{2} \int \rho \, \frac{ \partial \xi_i^*}{\partial t}   \frac{ \partial \xi^i }{\partial t}  dV 
= \frac{1}{2} \sum_n \mathcal A_n^2 \, |\dot a_n(t)|^2   \, ,   \label{eq:Ek} \\
 E_p (t) &  = \frac{1}{2} \int ( \xi_i^*  C \xi^i ) \,  dV 
= \frac{1}{2} \sum_n \mathcal A_n^2 \, \omega_n^2 |a_n(t)|^2  \, ,  \label{eq:Ep}
\end{align}
where we have used
\begin{equation}
\langle \xi _{n}, C \xi_n \rangle = \omega_n^2 \langle \xi _{n}, \rho \xi_n \rangle = \mathcal A_n^2 \, \omega_n^2 \, .
\end{equation}
The total tidal energy is therefore given by
\begin{equation}
E = E_k  + E_p = \frac{1}{2} \sum_n \mathcal A_n^2 \, \left( |\dot a_n(t)|^2  + \omega_n^2 |a_n(t)|^2  \right) \,
\end{equation}
and the rate of energy transfer from the tide to the oscillation modes is obtained from  \citep{l94}
\begin{equation}
\dot E  = - \int \rho \frac{ \partial \xi^i}{\partial t}    \nabla \chi_i^* \,  dV
=  \sum_n  \left( G M' \frac{W_{lm} Q_n}{D^{l+1}} e^{i m \Phi (t)} \right) \, \dot a_n(t) \label{eq:dEdt} \, .
\end{equation}
Differently from \citet{l94}, we consider only the energy of a single mode and not the pair of $m=\pm 2$ modes (for $l=2$). 
The maximum mode energy just after the resonance can be estimated as
\begin{equation}
 \tilde E_{\rm max} \simeq \frac{\pi^2}{ 1024 } \, \frac{\tilde Q_n ^2}{ \tilde {\mathcal A}_n^2} \, \, \tilde \omega_n ^{1/3} \left( \frac{R c^2}{ G M_\star } \right)^{5/2} q \left( \frac{2}{1+q} \right)^{5/3} \, . \label{eq:Emax}
\end{equation}
In practice, we
 find that equation (\ref{eq:Emax}) provides values which are about 25\% smaller than the maximum energy determined from numerical solutions \citep[see][for a similar result]{wein}.

%]}
%-------------------FIG. 9--------------------------%
\begin{figure*}
\begin{center}
\includegraphics[height=70mm]{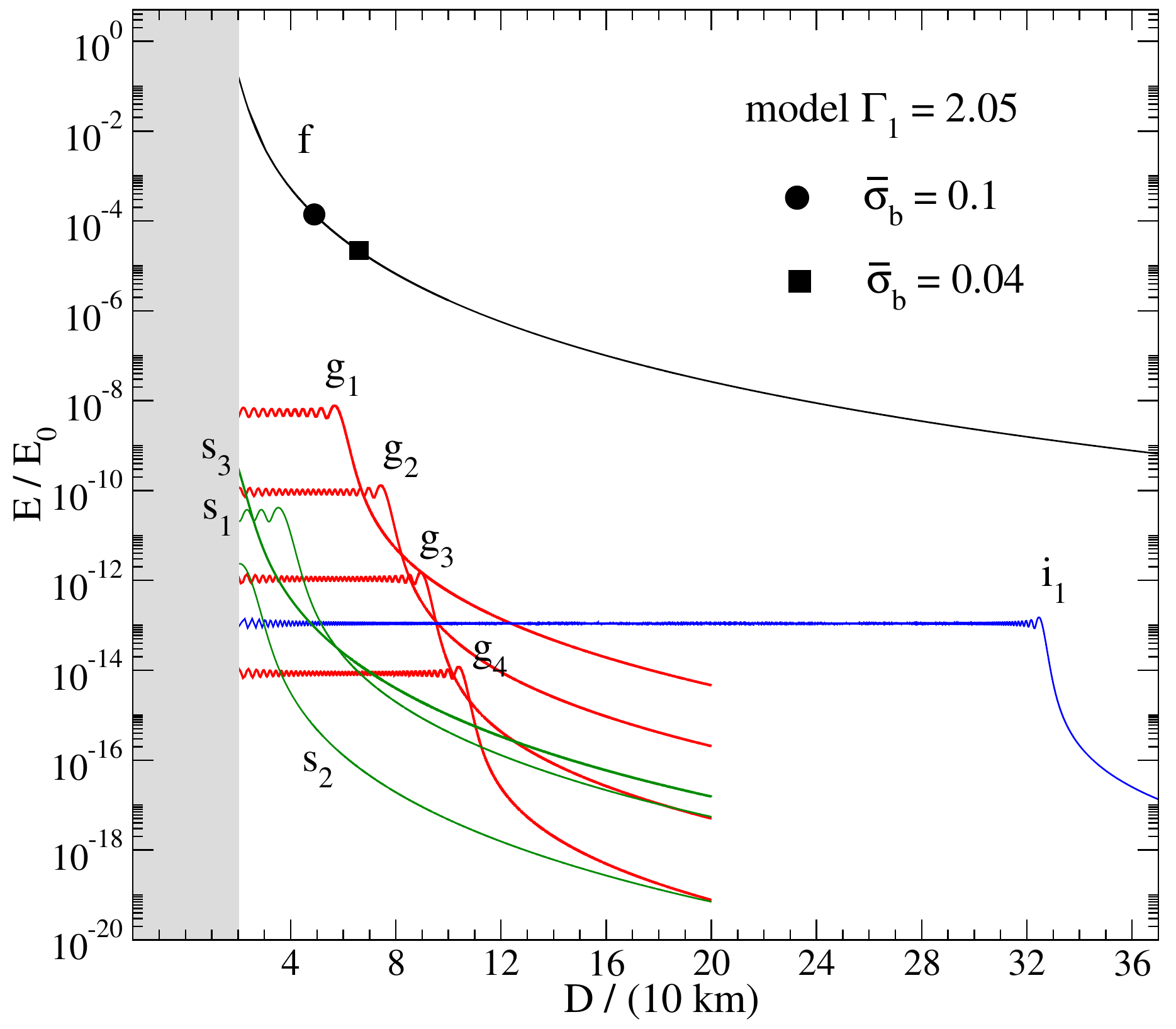} 
\includegraphics[height=70mm]{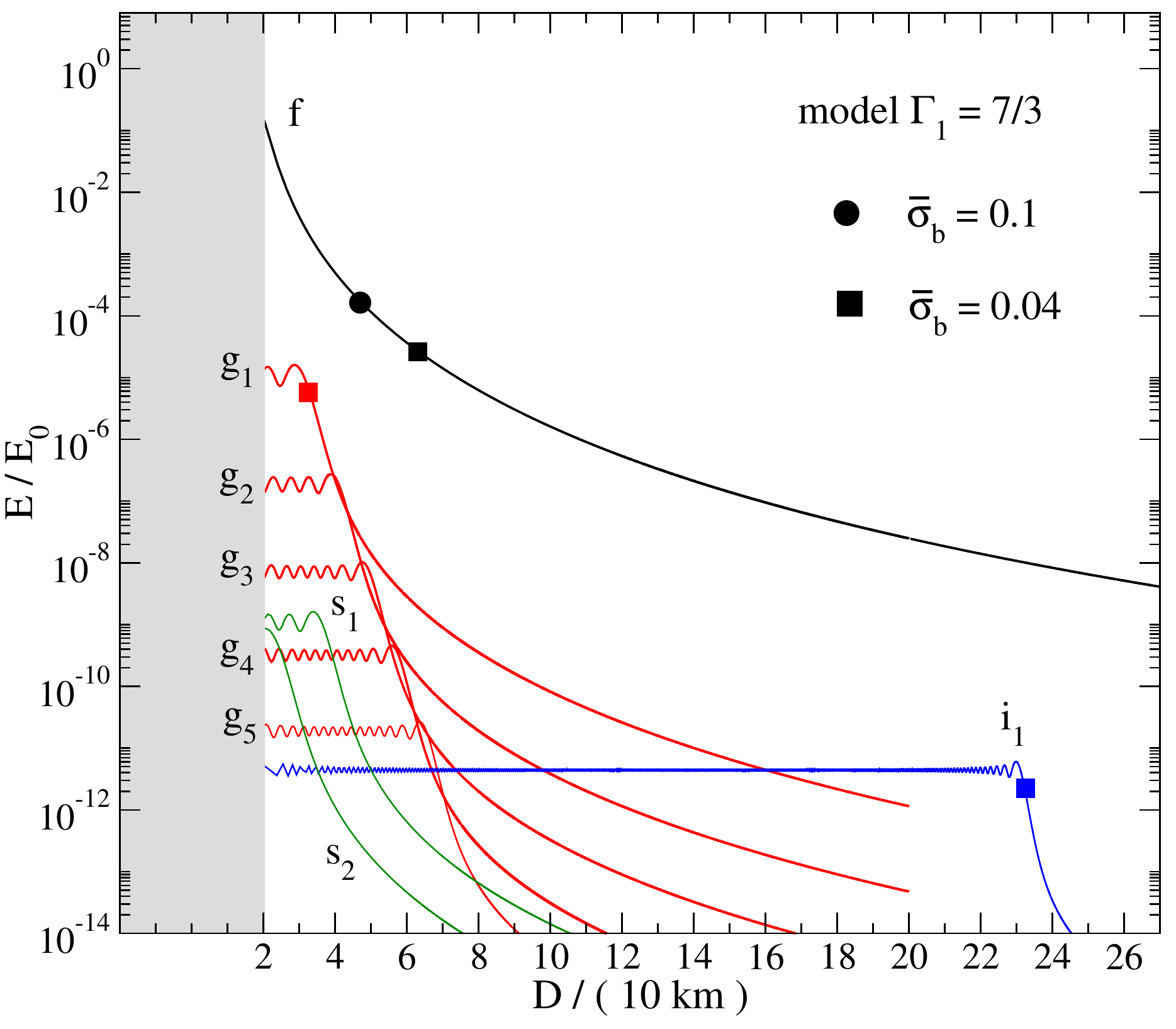} 
\caption{The evolution of the mode energies for varying binary separation. The energy is given in units of $E_0 = G M_\star^2 /R$. The model with $\Gamma_1=2.05$ is shown in the left-hand panel, while the $\Gamma_1=7/3$ case is provided in the right-hand panel. The energy of a mode increases when it becomes resonant during the inspiral. The breaking energy limit $E_b$ of the crust is represented by 
a circle or a square (see legend), which, respectively, denote the breaking energy for two choices of the breaking strain, $\bar \sigma_b = 0.1$ and $\bar \sigma_b = 0.04$. 
 \label{fig:res}}
\end{center}
\end{figure*}
%-----------------------------------------------------%

We have quantified the mode resonances for an equal-mass binary system with $M_\star= 1.4 M_{\odot}$, and $R=10$ km. 
In Figure \ref{fig:res} we show the evolution of the mode energy during the inspiral for the two stratified models with
$\Gamma_1 = 2.05$ and $\Gamma_1 = 7/3$. The results show how the
various oscillation modes are resonantly excited as the binary separation decreases towards merger, roughly corresponding to $D = 2 R$.
The first modes to be excited are low-frequency modes, like the surface g modes (not shown in the figure), the interface modes and high-overtone g modes. 
As the inspiral proceeds, the orbital angular velocity increases and low-order g modes and shear modes are progressively satisfying 
the resonant condition. However, it is clear from the results that the f mode always dominates the dynamical tide, even though it does not become resonant until after merger (in this example).
The energy of other resonant modes does not at any point reach above about 1\% of the f-mode energy. This accords well with the discussion of \citet{ap20a}, and supports the assumptions made by \citet{ap20b}.
The maximum g-mode energy grows larger as the stratification becomes stronger, but the resonance occurs at a later time, closer to the merger. 
It is worth noting that, after the resonance, the modes keep oscillating due to the absence of dissipative processes in our model.
This effect would be (at least to some extent) suppressed if we were to include viscous damping \citep[see][]{l94}. 

\begin{figure}
\begin{center}
\includegraphics[height=70mm]{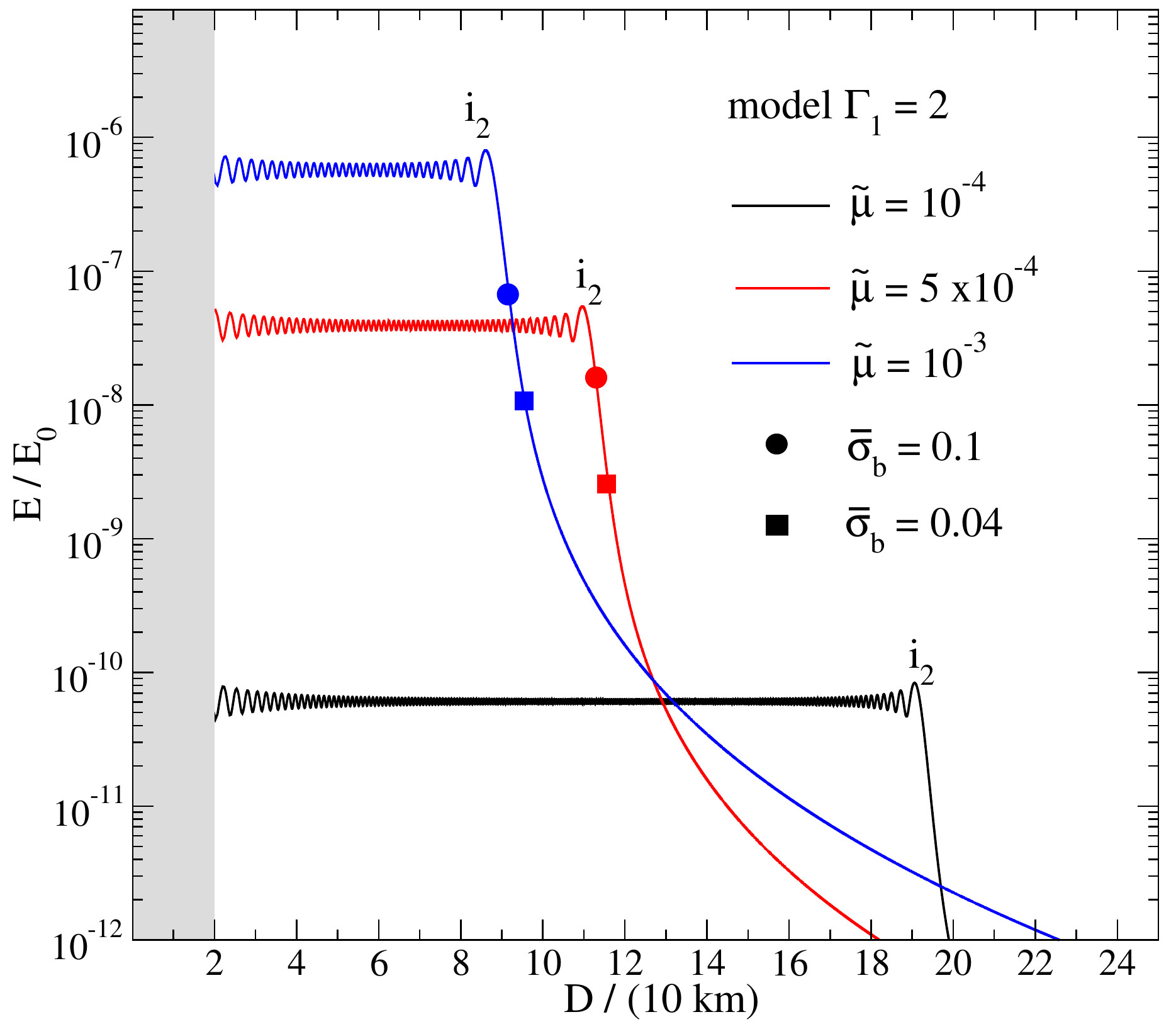} 
\caption{Illustrating the dependence of the core-crust interface (i$_2$) mode resonance excitation on the shear modulus of the crust. The results are for the  barotropic model ($\Gamma_1=2$). 
The three curves show the results for three different values of the shear modulus: $\tilde \mu = 10^{-4}$, $5 \times 10^{-4}$ and $10^{-3}$ (see legend). The shear modulus is given in units of $G M_\star /R$ and the energy in units of $E_0 = G M_\star^2 /R$. The breaking energy limit $E_b$ of the crust is represented by a circle or a square (see legend), which, respectively, denote the breaking energy for two choices of the breaking strain, $\bar \sigma_b = 0.1$ and $\bar \sigma_b = 0.04$. 
 \label{fig:res-imode}}
\end{center}
\end{figure}

For the i$_1$ interface mode the maximum energy is very small, although it increases with stratification.  It is about an order of magnitude larger for the $\Gamma_1 = 7/3$ model compared to the $\Gamma_1=2.05$ case. 
Having a lower frequency compared to the other modes (see  Figure~\ref{fig:res}) the interface mode enters resonance earlier, when the orbital separation is  
about  320 km for the model with $\Gamma_1 = 2.05$ and 230 km for the model with stronger stratification. 
In this earlier phase, the binary evolution is slower and as a result the interface modes have more time to accumulate energy. This will be important when we consider the issue of crust failure. Before we consider this question, it is useful to assess the impact of the shear modulus on the interface modes. 
For the barotropic model  ($\Gamma=2$), we determine the i$_2$  mode---which originates at the 
crust-core interface---for various values of the shear modulus parameter $\tilde \mu$, respectively,  
$\tilde \mu = 10^{-4}$, $5 \times 10^{-4}$ and $10^{-3}$ (see Equation \eqref{eq:mu}).  The results are shown in 
Figure \ref{fig:res-imode}, from which it is evident that that the i$_2$ mode  depends strongly on the crust rigidity.

%---------------------FIG. 10-------------------------%
%-----------------------------------------------------%

%%%%%%%%%%%%%%%
\subsection{Breaking the crust}  \label{sec:Eb}

Let us now quantify the resonant mode amplitudes relative to the level required to exceed the breaking strain of the crust.
The mode energy is not enough to answer this question, we also need to evaluate the elastic strain associated with the mode and this depends on the detailed eigenfunctions.  

The crust problem is complex (including aspects that are difficult to pin down, like the impact of possible pasta regions close to the core-crust transition), but we can make progress by combining the mode eigenfunctions, an estimate for the breaking strain and the standard von Mises criterion.  
We first of all need the elastic strain tensor. Hence, we define the tensor field $\bar  \sigma_{ij} =  \sigma_{ij} / \check \mu$ (not to be confused with the quantity used in the formal orthogonality analysis), 
such that
\begin{equation}
\bar   \sigma_{ij} =  \left( \nabla_{i} \xi_{j} + \nabla_{j}
    \xi_{i} \right) - \frac{2}{3}  g_{ij}  \left(
    \nabla_{k}\xi^{k} \right) \,  .
\end{equation}
Through the von Mises criterion, the crust breaking is then established by comparing
\begin{equation}
\bar \sigma \equiv \sqrt{ \frac{1}{2} \bar \sigma_{ij}^* \bar \sigma^{ij}} \, . \label{eq:sig_b}
\end{equation}
to the breaking strain. Based on the molecural dynamics simulations of \citet{horo} the breaking strain is commonly taken to be  $\bar \sigma_b = 0.1$, so we naturally focus on this case.
 At the same time it is useful to ask how the results depend on the this assumption. Hence, we also consider the results from \citet{baiko} which suggest the slightly lower value of  $\bar \sigma_b = 0.04$. 

Introducing the vector expansion of the Lagrangian displacement (\ref{eq:xiexp}) into equation (\ref{eq:sig_b}) we obtain the following expression for an $l=m=2$ mode: 
\begin{equation}
\bar \sigma^2 = \frac{5}{\pi} \frac{\sin^4 \theta}{r^2}  \left[ \frac{1}{4} \left(\frac{ dW}{dr} \right) ^2 
- \frac{3}{4} \left(\frac{ dV}{dr} \right) ^2 
+ \left( \frac{3}{2} V - W \right) \frac{1}{r} \frac{ dW}{dr}
+  \left(  V - \frac{W}{2} \right) \frac{3}{r} \frac{ dV}{dr}
+  \frac{1}{4} \left(  \frac{W}{r} \right)^2 \right] 
\end{equation}
From this expression we can then determine, for each mode, the oscillation energy required to satisfy the $\bar \sigma = \bar \sigma_b$ condition at some point in the crust. 
For each mode we have considered, and the three stellar models, we report in Tables \ref{tab:Eb25}, \ref{tab:Eb73} and \ref{tab:Eb2} 
 the breaking energy in the crust, $E_b$, and the maximum energy reached during an inspiral. 
 To determine the energy for the case $\bar \sigma_b = 0.04$, one can easily rescale the results given for $\bar \sigma_b = 0.1$ by using 
$E_b |_{\bar \sigma = 0.04} = 0.16 \, E_b |_{\bar \sigma = 0.1}$.  

%------------------------------FIG. 11--------------%
\begin{figure*}
\begin{center}
\includegraphics[height=45mm]{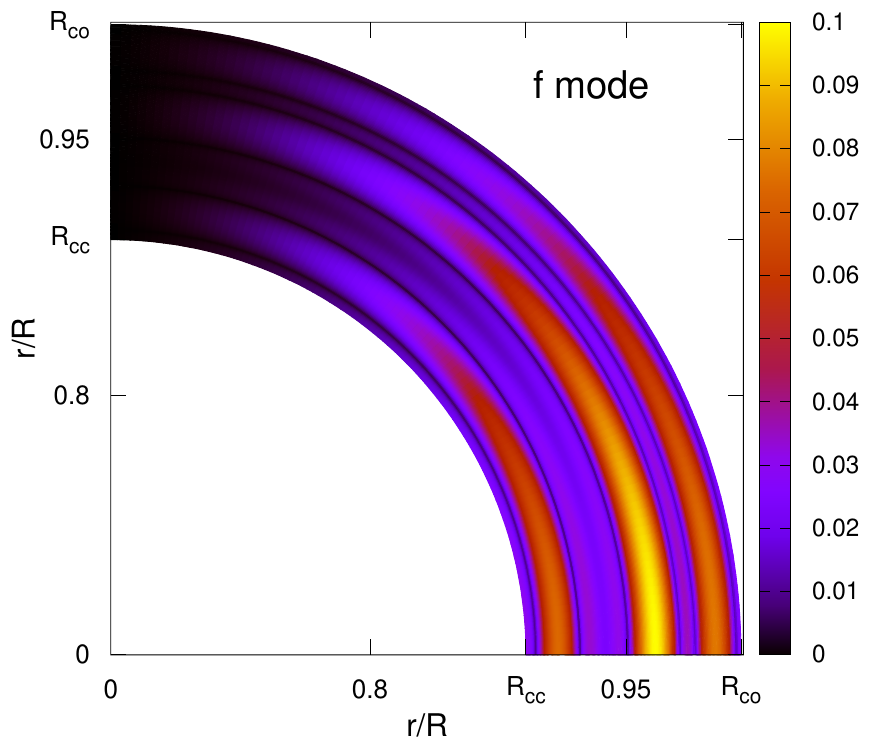} 
\includegraphics[height=45mm]{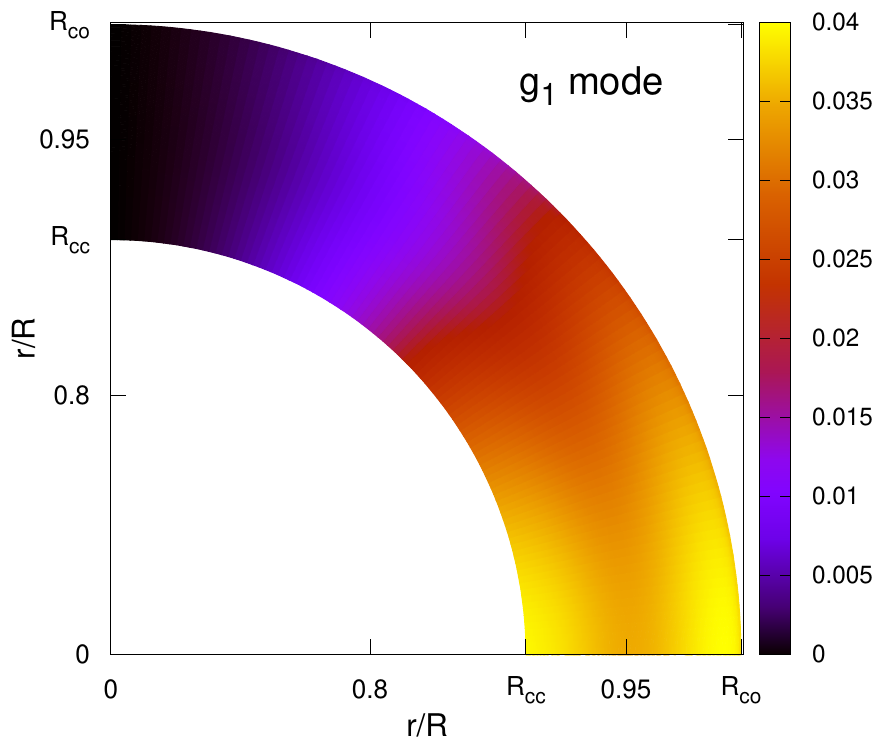} 
\includegraphics[height=45mm]{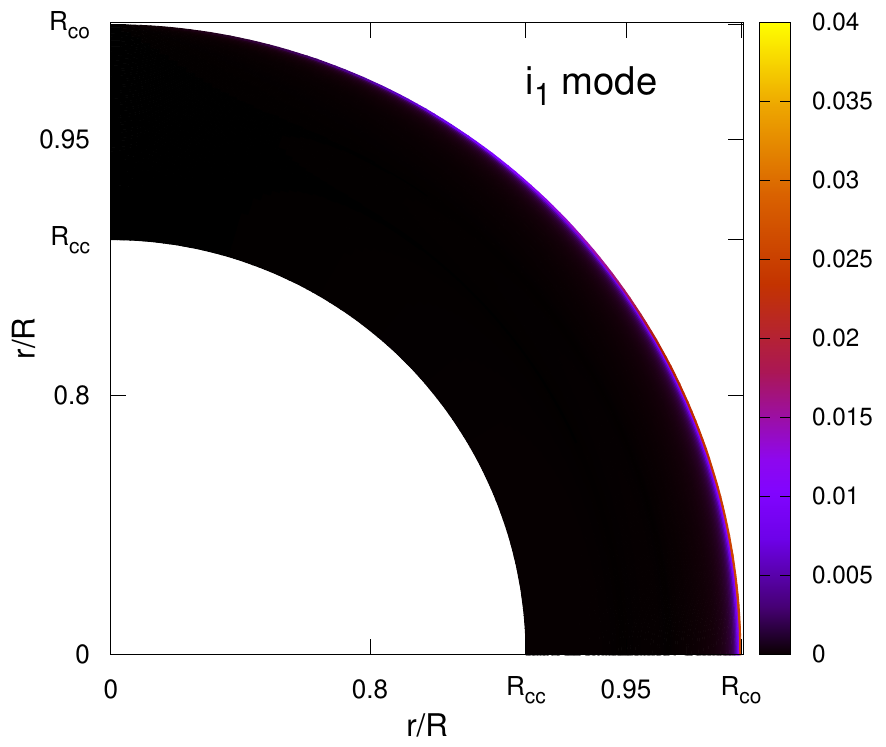} 
\caption{The strain field $\bar \sigma$ for the resonant modes which reach the  breaking limit, $\bar \sigma_b$, during inspiral. 
For the stratified model with $\Gamma_1=7/3$, we show, from the left to the right panel, meridional 2D cross sections for the f, g$_1$ and i$_1$ modes. 
The f mode can reach the breaking limit $\bar \sigma_b=0.1$, while the  g$_1$ and i$_1$ modes can fracture the crust only if we consider the lower breaking limit 
$\bar \sigma_b=0.04$ (see the bar legend).  Lighter colours indicate larger strain.  
 \label{fig:sigma1}}
\end{center}
\end{figure*}
%-----------------------------------------------------------%
%--------------------TAB. 0-------------------------------%
\begin{table}
\begin{center}
\caption{\label{tab:Eb25} Mode excitation and breaking energy for the stratified model  with $\Gamma_1=2.05$. We provide the maximum resonant energy $E_\mathrm{max}$ (second column) and  the breaking energy $E_b$  (third column; normalised to $E_0 = G M_\star^2 /R$) for each mode (first column) we have considered. The breaking energy is determined from the von Mises criterion for $\bar \sigma_b = 0.1$}
\begin{tabular}{c  c c   }
\hline
  Mode & $   E_{\rm max} / E_0 $  &  $ E_{b} / E_0 $  \\ % &  $ r_b /R $  \\
  &    &  $ \bar \sigma_{b} =0.1 $  \\  
\hline
f	& 	$4.28$ &		                          $ 1.39\times 10^{- 4 } $	   \\ % & 0.963 \\
g$_1$	& 	$7.75\times 10^{-9  }   $    &      $ 6.13\times 10^{-6  } $\\ %  &	0.900 \\
g$_2$ 	 & 	$1.36\times 10^{- 10 } $  &	$ 1.62\times 10^{-6  } $ \\ % &	0.900 \\
g$_3$	&  	$1.72\times 10^{- 12 } $  &	$8.91\times 10^{- 7 } $ \\ % & 0.900 \\
g$_4$	&  	$2.31\times 10^{-  14} $	 &	$7.88\times 10^{- 7 } $ \\ % & 0.900 \\
g$_{1}^s$	 & 	$6.53\times 10^{-  17} $   &	$6.12\times 10^{-11  } $ \\ %& 0.999 \\
g$_{2}^s$	& 	$4.75\times 10^{-  18} $  &	$1.63\times 10^{- 10 } $\\ % & 0.999 \\
g$_{3}^s$	 & 	$2.41\times 10^{-  19} $ &		$5.01\times 10^{-  10} $ \\ %& 0.999 \\
i$_1$ 	& 	$1.51\times 10^{-  13} $ &		$3.32\times 10^{-  12} $ \\ %& 0.999 \\
s$_1$	 & 	$4.24\times 10^{-  11} $  &	$1.45\times 10^{- 8 } $ \\ %& 0.951 \\
s$_2$	 & 	$2.62\times 10^{-  12} $ &		$8.05\times 10^{- 9 } $ \\ %& 0.973 \\
s$_3$	 & 	$1.34\times 10^{-  9}   $  &	$5.61\times 10^{- 9 } $ \\ %& 0.981 \\
%$s_4$	 & 	      	 				&	$6.99 \times 10^{-  5} $ \\ %& 0.987\\
%$s_5$	 &						&	 $3.52\times 10^{- 9 } $ \\ %& 0.988\\
%$s_6$	 & 						&	$ 2.97 \times 10^{-  9} $\\ % & 0.990\\
\hline   
\end{tabular}
\end{center}
\end{table}
%------------------------------------------------------------------------------%

%------------------------------TAB. 0------------------------------------------%
\begin{table}
\begin{center}
\caption{\label{tab:Eb73} The same as Table \ref{tab:Eb25}, but for the model with $\Gamma_1=7/3$. }
\begin{tabular}{c  c c   }
\hline
  Mode & $   E_{\rm max} / E_0 $  &  $ E_{b} / E_0 $  \\ % &  $ r_b /R $  \\
&    &  $ \bar \sigma_{b} =0.1 $  \\  
\hline
f	         & 	$4.04$                              &      $ 1.63\times 10^{- 4 } $	   \\ % & 0.963 \\
g$_1$	& 	$1.63\times 10^{-5  }   $    &      $ 3.59\times 10^{-5  } $\\ %  &	0.900 \\
g$_2$ 	 & 	$2.72\times 10^{- 7 } $  &	$ 1.96\times 10^{-6 } $ \\ % &	0.900 \\
g$_3$	&  	$1.02\times 10^{- 8 } $  &	$3.71\times 10^{- 6 } $ \\ % & 0.900 \\
g$_4$	&  	$4.85\times 10^{-  10} $	 &	$3.19 \times 10^{- 6 } $ \\ % & 0.900 \\
g$_5$	&  	$2.71\times 10^{-  11} $	 &	$3.24\times 10^{- 6 } $ \\ % & 0.900 \\
g$_{1}^s$	 & 	$4.10\times 10^{-  14} $   &	$8.39\times 10^{-11  } $ \\ %& 0.999 \\
g$_{2}^s$	& 	$3.78\times 10^{-  15} $  &	$2.05\times 10^{- 10 } $\\ % & 0.999 \\
g$_{3}^s$	 & 	$6.41 \times 10^{- 23} $ &		$6.22\times 10^{-  6} $ \\ %& 0.999 \\
i$_1$ 	& 	$6.06 \times 10^{-  12} $ &		$1.24\times 10^{-  11} $ \\ %& 0.999 \\
s$_1$	 & 	$1.63\times 10^{-  9} $  &	$1.65\times 10^{- 8 } $ \\ %& 0.951 \\
s$_2$	 & 	$9.06\times 10^{-  10} $ &		$8.16\times 10^{- 9 } $ \\ %& 0.973 \\
s$_3$	 & 	$1.68\times 10^{-  8}   $  &	$5.64\times 10^{- 9 } $ \\ %& 0.981 \\
%$s_4$	 & 	      	 				&	$6.99 \times 10^{-  5} $ \\ %& 0.987\\
%$s_5$	 &						&	 $3.52\times 10^{- 9 } $ \\ %& 0.988\\
%$s_6$	 & 						&	$ 2.97 \times 10^{-  9} $\\ % & 0.990\\
\hline   
\end{tabular}
\end{center}
\end{table}
%------------------------------------------------------------------------------%

%------------------------------TAB. 0------------------------------------------%
\begin{table}
\begin{center}
\caption{\label{tab:Eb2} Mode excitation and breaking energy for the barotropic model ($\Gamma_1=2$) and the i$_2$ interface mode.  We provide the maximum resonant energy $E_\mathrm{max}$ (second column) and  the breaking energy $E_b$  (third column; normalised to $E_0 = G M_\star^2 /R$) for three different values of the shear modulus parameter $\check \mu$ (first column). The breaking energy is determined from the von Mises criterion for $\bar \sigma_b = 0.1$.}
\begin{tabular}{c  c c   }
\hline
    $\check \mu$     &     $   E_{\rm max} / E_0 $   &  $ E_{b} / E_0 $  \\         &  &  $ \bar \sigma_{b} =0.1 $  \\  
\hline
  $10^{-4} $	 & $8.36 \times 10^{-  11} $ &		$3.99\times 10^{-  9} $ \\
 $5 \times 10^{-4} $	 & 	$5.47 \times 10^{-  8} $ &		$4.35\times 10^{-  8} $ \\
 $10^{-3} $	 & 	$8.00 \times 10^{-  7} $ &		$1.72\times 10^{-  7} $ \\
\hline   
\end{tabular}
\end{center}
\end{table}
%------------------------------------------------------------------------------%

The results for the crust fracturing show, first of all, that the f mode (which dominates the tidal response, see Figure \ref{fig:res}) may break the crust when the system is close to merger, roughly at a separation of  $D \simeq 50-70$ km (depending on the chosen value for $\bar \sigma_b$). For the weakly stratified model, we find that the interface i$_1$ mode reaches an energy slightly lower than the breaking limit. In contrast, 
for the strongly stratified model, the g$_1$ and i$_1$ modes both reach an energy above $E_b$ and hence may impact on the crust. The former mode reaches the breaking amplitude during the late stages of the inspiral, even later than the f mode. Perhaps more interesting, in this respect, is the i$_1$ mode which may break the crust at a much larger separation,  $D \simeq 230$ km. 
As discussed in Section~\ref{sec:modes}, the g modes and the core-crust interface mode have similar eigenfunctions in the stratified models. To study the impact of the i$_2$ mode on the crust, we therefore consider the barotropic star and explore the effect of the shear modulus on the mode dynamics. For the three values of the shear modulus considered in Section~\ref{sec:dyn}, the i$_2$ mode overcomes the breaking energy $E_b$ only in models with $\tilde \mu \ge 5 \times 10^{-4}$. 

Crust failure due to the resonance of the i$_2$ mode has been previously studied by \citet{tsang} and \citet{bonga}. In general, we find that the maximum i$_2$ mode energy is close to the crust breaking limit, but the crust only fractures for some of the models. In our models the overlap integral for the i$_2$ mode is about two orders of magnitude smaller than those reported in \citet{tsang} and \citet{bonga}. This difference is likely due to the different equation of state and  shear modulus prescription.  \citet{tsang} and \citet{bonga} use the Newtonian perturbation equations, as in our work,  but  the stellar model is described by tabulated equations of state. 
 In particular, in  \citet{tsang} the equilibrium star is determined by using the relativistic structure equations. 
As the i$_2$ mode eigenfunctions depend on the properties of the stellar models, as we have  demonstrated,  it is not surprising that we  arrive at slightly different results. The main implication is clear. If we want to draw firm conclusions on the likelihood of crust failure due to the interface mode, we need to use a more realistic model and a complete relativistic formulation of the problem.

Turning to the question of the location at which the crust first fails, we note that the largest stress is reached at the equator for all the modes we consider.
The exact location where a mode reaches the breaking limit is indicated in 
Figure \ref{fig:sigma1}. The f mode first reaches the threshold for crust failure at $r\simeq0.96 R$, in the low density region close to the surface. In contrast,  the g$_1$ mode 
stresses the crust predominantly at the crust-core and crust-ocean interfaces, 
  while the i$_1$ mode  strain reaches its maximum value at the crust-ocean transition, as anticipated given the distinct cusp at the transition density. 
For the barotropic model $(\Gamma_1=2)$,  the strain tensor $\bar \sigma_{ij}$ for the i$_2$ mode has a larger magnitude at the top of the crust but reaches a significant level throughout the equatorial plane. This property is more pronounced in the $\tilde \mu = 10^{-3}$ case, and may be significant as it could indicate a global, rather than local, crust failure.

\section{Concluding remarks}

We have studied the tidal response of a binary neutron star during the inspiral phase, considering spherically symmetric models with crust and ocean. 
First, we  revisited the theoretical formulation of the problem to understand whether the presence of a crust or density discontinuities 
require  changes to the formalism so far developed.  Our analysis shows that these new ingredients do not change substantially 
the Newtonian formalism used for fluid  models as long as the fundamental boundary and junction conditions, at the different interfaces, are satisfied.  

Considering the various oscillation modes sustained by our stellar models, we have extended the previous analysis of \citet{ap20a}
focusing on the effect of the crust. In particular we have studied the Love number in the static limit, considering the contribution from the most relevant 
oscillation modes.  We have shown that the presence of the crust does not significantly affect  the fundamental, pressure and gravity modes. 
As expected, the contribution to the Love number from the shear modes is negligible, while 
 the interface and surface gravity modes have an impact similar to the first core gravity modes. 
 The influence of these modes, albeit small compared to the fundamental mode,  
 increases for strongly stratified models.

Oscillations may be  amplified by tidal resonances during the binary inspiral. 
This amplification is not only important for the gravitational wave signal, but also for the impact that mode resonances can have on the crust.  
We have studied the dynamical tidal evolution and determined the mode energy during the orbital shrinking. 
Our results confirm that the fundamental mode dominates the dynamical tides even when it is far from resonance. In our models, the f mode would enter resonance 
for a  binary separation   $D/R \simeq 1.74$, i.e. after merger. 
Among the other modes, the first gravity mode reaches the largest oscillation energy during the late stages of the inspiral, roughly when
$D/R \simeq 5-10$ (depending on the stellar model). 
The interface modes are resonantly excited at an earlier stage and may accumulate enough energy to fracture the crust. 
This is mainly due to their peaked radial displacement at the crust-core or crust-ocean surface transition. 
We have used the von Mises criterion to determine the minimum energy required to fracture the crust and compare the result to  the energy 
gained by a given mode during the binary evolution. The interface modes do not break the crust for all our models. Strongly stratified cases 
are favoured in this respect. This is not surprising, because it is well known that the interface mode properties depend on the equation of state, shear modulus, 
density discontinuities, etcetera. A variation in these quantities can lead to different conclusions. 
We also found that the interface modes mainly break the crust at the equator and predominantly at the crust-ocean transition. 

The fundamental mode reaches the crust breaking limit in all our models, but not until the final part of the inspiral. The f-mode eigenfunctions 
have a more regular behaviour at the crust boundaries than the interface modes, therefore it needs to reach a larger energy in order to 
fracture the crust.  We find that the strain tensor for the f mode reaches its largest value around the middle of the crust. 
Finally, we have shown that the first gravity mode can fracture the crust only for strongly stratified stars and (again) in the very final phase of inspiral.  

A natural extension of this work would be the inclusion of superfluid and superconducting constituents in the core and the inner crust. For such models we expect that shear and gravity modes will be shifted towards higher frequencies \citep[see for instance][]{pass, wein}, but the problem is complicated as superfluid entrainment comes into play (and may have a particularly large effect in the crust).
The corresponding mode resonances would then be expected very close to the merger and might have negligible impact on the tidal problem. However, in order to quantify the effect, we need to add the superfluid degree of freedom to the analysis. We have already worked through the formal aspects of this and expect to complete the analysis with a sample of numerical results before too long.  
Another important step would be the development of a relativistic formulation of the problem.  This would allow us 
to study the crucial influence of realistic/tabulated equations of state and relativistic effects 
on the problem.

%------------------------------------------------------------------------------%

\section*{Acknowledgements}

N.A. and P.P. acknowledge support from the Science and Technology Facilities Council (STFC) via grant ST/R00045X/1. P.P. acknowledges support from the ``Ministero dell'istruzione, dell'università e della ricerca" (MIUR) PRIN 2017 programme (CUP:~B88D19001440001) and from the Amaldi Research Center funded by the MIUR programme ``Dipartimento di Eccellenza" (CUP:~B81I18001170001).

\section*{Data availability}

Additional data related to this article will be shared on reasonable request to 
the corresponding author.

\newpage
\appendix

%%%%%%%%%%%%%%%%%%%% SEC. %%%%%%%%%%%%%%%%%
\section{Numerical code} \label{sec:num}
%%%%%%%%%%%%%%%%%%%%%%%%%%%%%%%%%%%%%%%%%%%%%%%%%%%%%%

We determine the oscillation mode properties by solving the perturbation equations, obtained from the single-fluid limit of the  equations given by \citet{pass} for superfluid stars. The stellar model we consider has three regions:  core,  crust and  ocean. Therefore, we must impose junction conditions at the origin and the star's surface, as well as boundary/junction conditions at the crust-core and crust-ocean transitions. 

\begin{figure}
\begin{center}
\includegraphics[height=70mm]{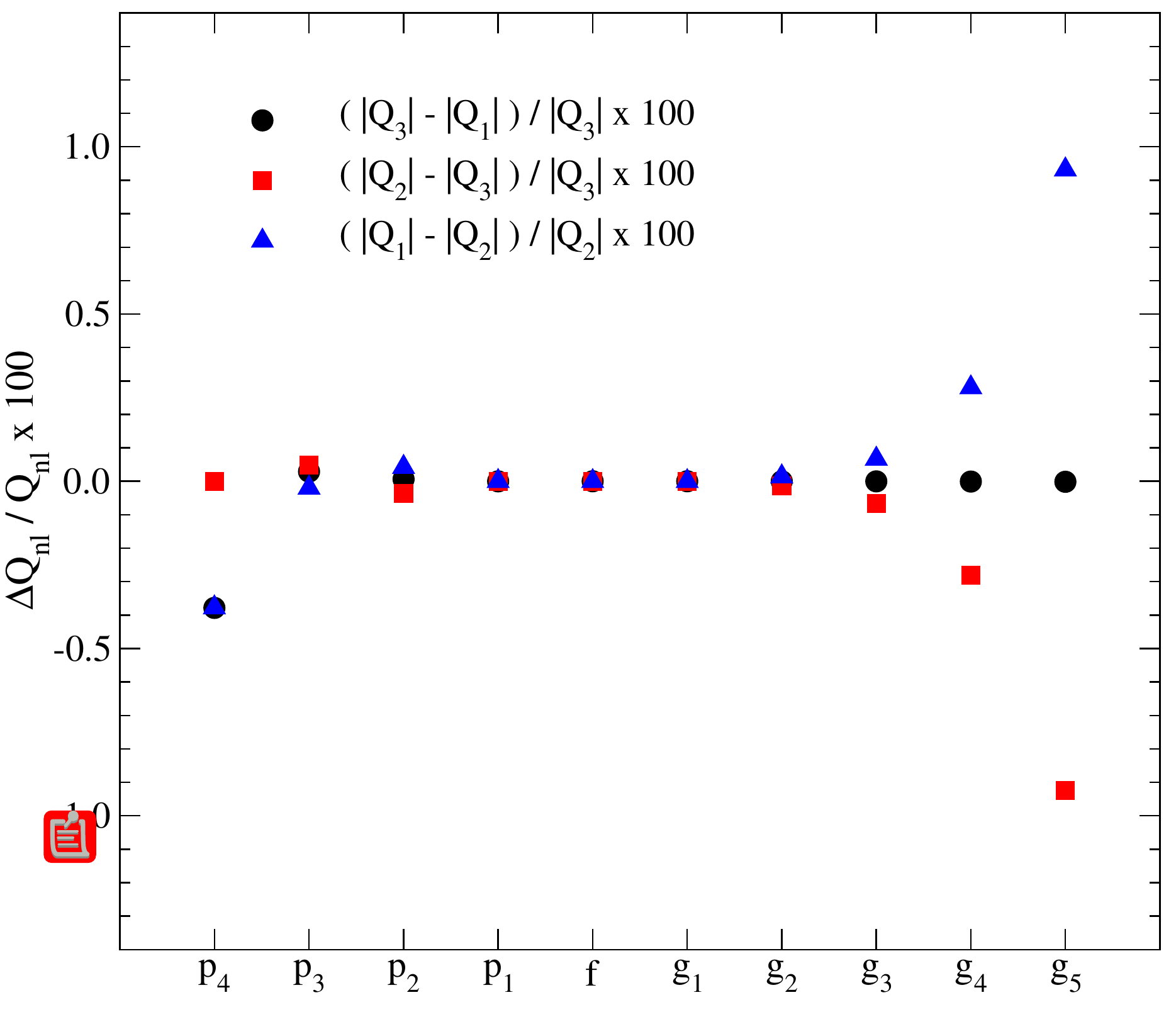} 
\includegraphics[height=70mm]{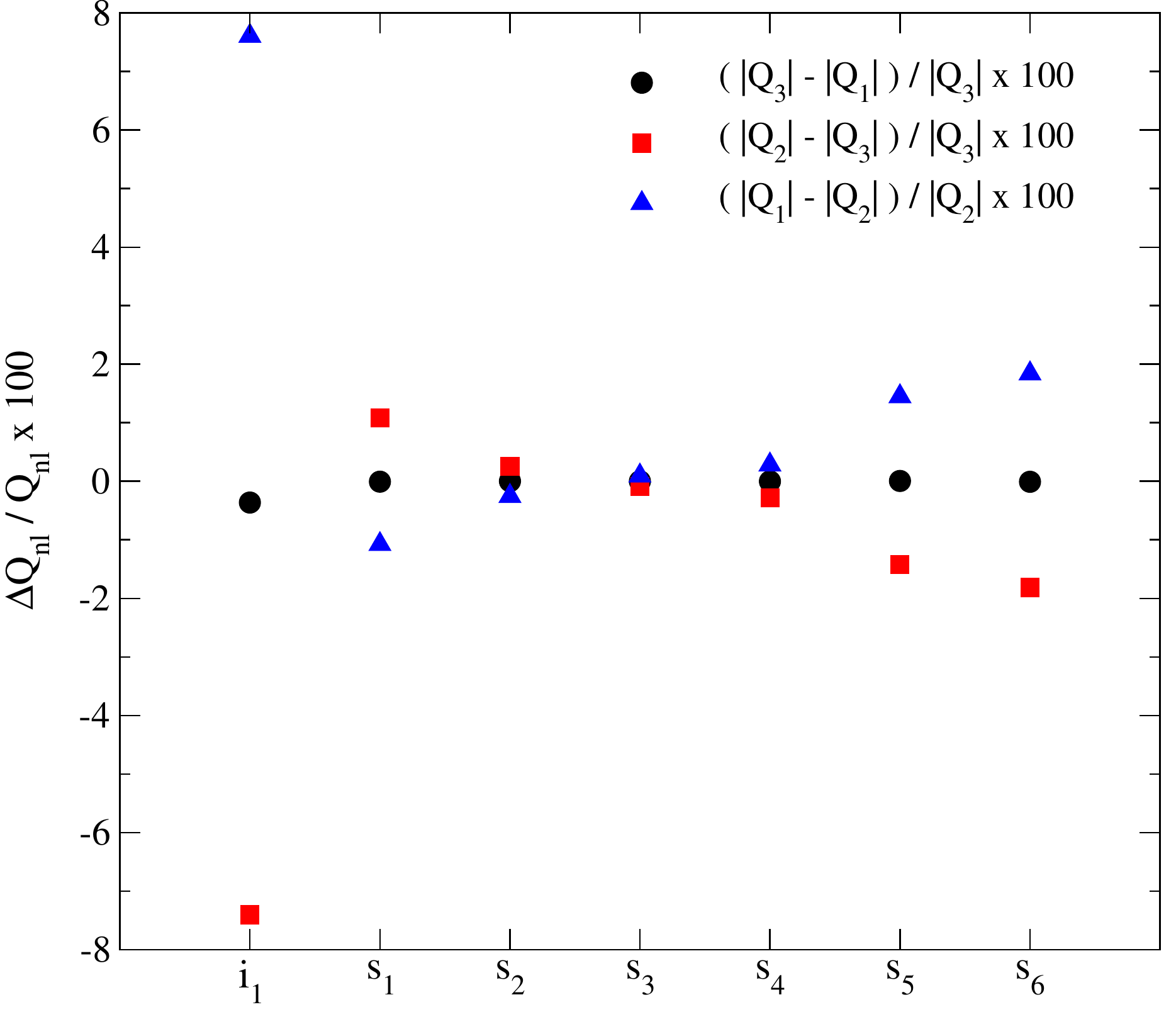} 
\caption{Percentage errors for the calculation of the overlap integral with the three expressions (\ref{overlap}), (\ref{multip}) and (\ref{eq:Qn_ter}), which are, respectively, denoted as $Q_1$, $Q_3$ and $Q_2$ in this figure. 
The results represent the stratified star with $\Gamma_1 = 7/3$.
The various oscillation modes are indicated on the horizontal axis. The overlap integral calculations agree to better than 1\%   for  most of the oscillation modes. 
The largest difference, $\lesssim 8\%$, is found for the interface modes. \label{fig:err}}
\end{center}
\end{figure}

We solve the linearised equations as an eigenvalue problem by using both multiple shooting methods and a relaxation approach. The latter was necessary to increase the accuracy of the calculated overlap integral. Basically, some oscillation modes have very small overlap integrals, the calculation of which may be subject to numerical errors. 
This is the case, for instance, for higher order modes which have many nodes in their eigenfunctions, and for shear and interface modes, which are mainly present in the crust region. 
To monitor the numerical errors we determine $Q_n$ from Equations (\ref{overlap}), (\ref{multip}) and (\ref{eq:Qn_ter}). 
The solutions obtained after the relaxation step   are much more accurate which allow us to study high-order oscillation modes. 
In Figure \ref{fig:err} we show the relative difference between the overlap integrals calculated from these three equations. 
The results agree to better than $1\%$ for all  modes, with the exception of the interface mode which can have an error at most of order a few~$\%$. To reach accurate results we have used a very high grid resolution with $1.92 \times 10^6$ points. 
All  other relevant quantities, as for instance mode frequencies and the eigenfunction ratio at the surface $(V/W)_R$, agree with the results from the literature also for lower resolutions.

%------------------FIG. 9------------------------------%
%---------------------%

%%%%%%%%%%%%%%%%%%%%%%% SEC. %%%%%%%%%%%%%%%%%%%%%%%%%%%%%%%%%%
\section{Mode resonance dynamics} \label{sec:mod-res}
%%%%%%%%%%%%%%%%%%%%%%%%%%%%%%%%%%%%%%%%%%%%%%%%%%%%%%%%%%%%%%%%%%%

In order to quantify the mode excitation during binary inspiral, we need to solve equation (\ref{eq:eq_a3}). We do this by introducing  
a new variable (this differs slightly from \citet{l94})  
\begin{equation}
a =  b \, e^{- i m \Phi(t)} \, .
\end{equation}
In terms of this new variable the amplitude equation takes the form
\begin{equation}
\ddot b_n - 2 i m \Omega \dot b_n  + \left( \omega_n ^2 - m^2 \Omega^2 - im \dot \Omega \right) b_n =    \frac{G M' }{R^3}  \frac{\tilde Q_n }{\tilde {\mathcal A}_n^2}  W_{lm} 
\left( \frac{R }{D(t)} \right)^{l+1}   \, .
\end{equation}
Introducing the decomposition $b = b^R + i b^I$ we obtain equations for the real and  imaginary parts, $b^R$ and $b^I$ respectively, of the scalar function $b$. These are
\begin{align}
& \ddot b_n^R + 2 m \Omega \dot b_n^I  + \left( \omega_n ^2 - m^2 \Omega^2 \right) b^R + m \dot \Omega b_n^I =    \frac{G M' }{R^3}  \frac{\tilde Q_n }{\tilde {\mathcal A}_n^2}  W_{lm} \left( \frac{R }{D(t)} \right)^{l+1}  \, , \label{eq:bR} \\
&  \ddot b_n^I - 2 m \Omega \dot b_n^R +  \left( \omega_n ^2 - m^2 \Omega^2 \right) b^I - m \dot \Omega b_n^R =    0  \, .
\label{eq:bI} 
\end{align}
Following \citet{l94}, we determine the initial conditions for $b^R$ and $b^I$ from the static limit of equation (\ref{eq:bR}), as the modes are then not resonant. By neglecting the time derivative in equation (\ref{eq:bR}) we obtain for the real part:
\begin{equation}
 b^R_0  =    \frac{G M^\prime }{R^3}  \frac{\tilde Q_l }{\tilde {\mathcal A}_n^2}  W_{nl} \left( \frac{R }{D(t_0)} \right)^{l+1}  \, \frac{1}{\omega_n ^2 - m^2 \Omega_0^2 } , \label{eq:bR_0} \\
\end{equation}
where quantities with the subscript ``0'' are  calculated at $t_0$. Inserting equation (\ref{eq:bR_0}) into  (\ref{eq:bI}), we have
\begin{equation}
\dot  b^R_0  \simeq  \left[ -(l+1) \frac{\dot D}{D} +    \frac{2 m^2 \Omega_0 \dot \Omega_0 }{\omega_n ^2 - m^2 \Omega_0^2 } \right] \,  b_0^R 
\simeq  - \left[ (l+1) +    \frac{3 m^2 \Omega_0 ^2 }{\omega_n ^2 - m^2 \Omega_0^2 } \right] \frac{\dot D}{D}  \,  b_0^R
\label{eq:dbRdt_0} \, , 
\end{equation}
where we have neglected $\ddot b_n^I$ in equation  (\ref{eq:bI}) and used
\begin{equation}
\frac{\dot \Omega}{\Omega } =  - \frac{3}{2} \frac{\dot D}{D}  \, .
\end{equation}
For the imaginary part we can choose 
\begin{align}
&  b^I _0 \simeq    \frac{m}{ \omega_n ^2 - m^2 \Omega^2} \left( 2 \Omega \dot b_0^R  +  \dot \Omega b_0^R \right) \, , \\
& \dot b^I _0 \simeq 0 \, .
\end{align}
This then allows us to solve the problem, leading to the results presented in Section~\ref{sec:frac}.

\end{document}